\documentclass[paper,12pt]{JHEP3}
\usepackage{epsfig}
\usepackage{amsmath}
\usepackage{amssymb}
\usepackage{epsf}
\usepackage{graphicx}

\title{Photon tagged correlations in heavy ion collisions}

\author{F. Arleo~$^{a,b}$, P. Aurenche~$^{c}$, Z. Belghobsi~$^{c,d}$, J.-Ph. Guillet~$^{c}$\\~\\
$^a$ Universit\'e de Li\`ege, Institut de
Physique B5,\\Sart Tilman, 4000 Li\`ege 1, Belgique\\~\\
$^b$ LPTHE~\footnote{UMR 7589 du CNRS}~, Universit\'e Paris VI {\cal \&} Paris VII et
CNRS,\\
4, Place Jussieu, 75252 Paris cedex 05, France\\~\\
$^c$ LAPTH~\footnote{UMR 5108 du CNRS, associ\'ee \`a l'Universit\'e de
Savoie}~, B.P.110, 74941 Annecy-le-Vieux cedex, France\\~\\
$^d$ Laboratoire de Physique Th\'eorique, Universit\'e de
Jijel,\\B.P. 98, Ouled Aissa, 18000 Jijel, Alg\'erie\\~\\
E-mail: \email{arleo@lpthe.jussieu.fr}, \email{aurenche@lapp.in2p3.fr}, \\\email{belghob@lapp.in2p3.fr}, \email{guillet@lapp.in2p3.fr}}


\preprint{JHEP {\bf 11} (2004) 009}

\newcommand{\slL}{\raise.15ex\hbox{$/$}\kern-.53em\hbox{$L$}}
\newcommand{\slP}{\raise.15ex\hbox{$/$}\kern-.67em\hbox{$P$}}
\newcommand{\slp}{\raise.1ex\hbox{$/$}\kern-.63em\hbox{$p$}}
\newcommand{\slq}{\raise.1ex\hbox{$/$}\kern-.63em\hbox{$q$}}
\newcommand{\slv}{\raise.1ex\hbox{$/$}\kern-.63em\hbox{$v$}}
\newcommand{\slR}{\raise.15ex\hbox{$/$}\kern-.53em\hbox{$R$}}
\newcommand{\slQ}{\raise.15ex\hbox{$/$}\kern-.53em\hbox{$Q$}}
\newcommand{\slK}{\raise.15ex\hbox{$/$}\kern-.53em\hbox{$K$}}
\newcommand{\slk}{\raise.15ex\hbox{$/$}\kern-.53em\hbox{$k$}}
\newcommand{\slSigma}{\raise.15ex\hbox{$/$}\kern-.53em\hbox{$\Sigma$}}
\newcommand{\slcalP}{\raise.15ex\hbox{$/$}\kern-.63em\hbox{$\cal P$}}
\newcommand{\slA}{\raise.15ex\hbox{$/$}\kern-.73em\hbox{$A$}}
\newcommand{\slbfA}{\raise.15ex\hbox{$/$}\kern-.73em\hbox{${\imb A}$}}
\newcommand{\slpartial}{\raise.15ex\hbox{$/$}\kern-.53em\hbox{$\partial$}}

\def\cO#1{{{\cal{O}}}\left(#1\right)} \def\pt{$p_{_T}$}
\def\ptpi{$p_{_{T_\pi}}$} \def\ptgamma{$p_{_{T_\gamma}}$}
\def\qt{$q_{_T}\ $} \def\kt{{k_{_T}}} \def\pt3{{p_{_T{_3}}}}
\def\pt4{{p_{_T{_4}}}} \def\gamgam{$\gamma-\gamma\ $}
\def\gampi{$\gamma-\pi^0\ $} \def\pipi{$\pi^0-\pi^0\ $}
\def\eps{\epsilon}

\newcommand{\be}{\begin{equation}} \newcommand{\ee}{\end{equation}}
\newcommand{\bea}{\begin{eqnarray}} \newcommand{\ena}{\end{eqnarray}}
\long\def\comment#1{ }    

\abstract{A detailed study of various two-particle correlation functions
involving photons and neutral pions is presented in proton-proton and
lead-lead collisions at the LHC energy. The aim is to use these
correlation functions to quantify the effect of the medium (in
lead-lead collisions) on the jet decay properties. The calculations
are carried out at the leading order in QCD but the next-to-leading
order corrections are also discussed. The competition between
different production mechanisms makes the connection between the jet
energy loss spectrum and the \gampi correlations somewhat indirect
while the \gamgam correlations have a  clearer relation to the jet
fragmentation properties.}

\begin{document}

\section{Introduction}

Electromagnetic probes have long been thought to be useful to detect
the formation of a quark-gluon plasma in ultrarelativistic heavy ion
collisions~\cite{wa98,indian,finnish,photonYR}.  Many observables
involving photons can in principle be used. One of the simplest, from
a theoretical point of view, is the single photon spectrum as a
function of the transverse momentum $p_{_T}$: it is expected that
secondary collisions in heavy ion scattering will produce an excess of
direct photons, as compared to proton-proton scattering, in an energy
domain a few times the plasma temperature.
However the flux of background photons from hadronic resonances is
quite large and this makes the extraction of the direct photon signal
a non trivial task experimentally. These observables would probe the
plasma at the early times of the collision when the medium is the
hottest.

Photons can also be used in a different kinematical regime, namely at
large transverse momentum (many times the temperature of the
plasma). Such photons are produced in primary collisions {\it i.e.} as
in proton-proton collisions and, ideally, their production rate is
calculable in perturbative QCD in the next-to-leading order (NLO)
approximation~\cite{hardphotons}. For a high enough transverse
momentum these direct photons should be easily extracted from the
background since the ratio $\gamma/\pi^0$ is rapidly increasing with
$p_{_T}$.  One can then study the decay properties of the jet
recoiling from the photon by considering various photon-hadron or
photon-photon correlation functions where the hadron or the second
photon are fragments of  the
jet~\cite{cfg1996,bgpw2001,diphox-program}. In the simplest case, when
the transverse momentum of the recoiling jet exactly balances that of
the photon, such observables allow to map out the fragmention function
of the jet traversing the medium~\cite{Wang:1996}. For a sample of
data large enough one hopes to study the difference in the shape of
the fragmentation functions in proton-proton and ion-ion
collisions. Of course the real situation is more complicated because
the large $p_{_T}$ photon can itself be produced by
bremsstrahlung~\cite{Sarcevic} in which case the photon and the jet
momenta become somewhat uncorrelated as it is also the case when
higher order corrections are taken into account.  Furthermore, in
order to have a reasonable counting rate for the correlation studies
one cannot consider photons with too large transverse momenta and the
$\pi^0$ background may then remain a problem.  Thus if one studies
\gampi correlations, the \pipi contribution  should also be considered
in turn.

In the following we will study various \gamgam, \gampi and \pipi
correlations both in proton-proton and lead-lead collisions at the
LHC.  Shadowing effects will be considered and, following standard
practice, we will assume that the effects on the hard process of the
parton multiple scattering through the medium can be parametrized by a
modification of the fragmentation functions. The results are obtained
in the leading-logarithm approximation of QCD since the status of NLO
calculations in a medium is not yet clear. Our results should
therefore be  considered only as semi-quantitative. The model will be
presented in the next section with special emphasis on the medium
modified fragmentation effects. Then we turn to several observables
and compare their behavior in proton-proton and lead-lead
collisions. A discussion of the effects of NLO corrections in
proton-proton collisions is given specifically to test the stability
of the shape of observables.  We consider this to be indicative of the
stability of correlations under higher order corrections in heavy ion
collisions.

\section{The model}

\subsection{The leading order cross section}

\begin{figure}[!ht]
\begin{center}
\includegraphics[height=5.5cm]{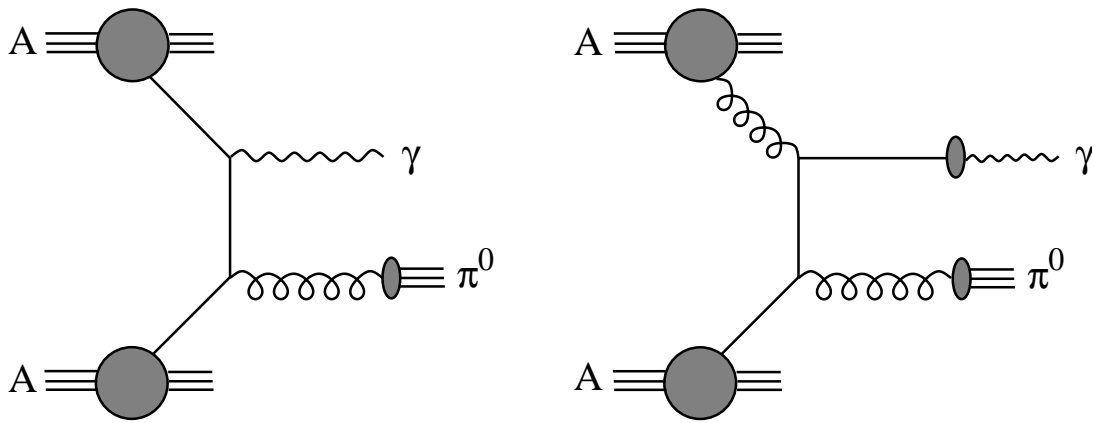}
\end{center}
\vspace{-0.1cm}
\caption{Two processes which contribute to \gampi production at
leading order: the photon may be produced directly (one-fragmentation,
labeled {\rm 1f} in the text, {\it left}) or by parton fragmentation
(two-fragmentation, labeled {\rm 2f}, {\it right}).}
\label{fig:1f2f}
\end{figure}

At the leading order in QCD the basic two-particle correlation
cross-section, from which we will construct various observables, can
be written~\cite{cfg1996}
\begin{eqnarray}
{d \sigma^{^{AB \rightarrow CD}} \over d p_{_T{_3}} dy_3 dz_3 d
p_{_T{_4}} dy_4 dz_4} = {1 \over 8 \pi s^2} & \sum_{a,b,c,d} &  {
D_{C/c}(z_3,M_{_F})\over z_3 } { D_{D/d}(z_4,M_{_F})\over z_4 }\
k_{_T{_3}} \  \delta(k_{_T{_3}}-k_{_T{_4}}) \nonumber\\  &&
{F_{{a/A}}(x_1,M) \over x_1} \ { F_{{b/B}}(x_2,M) \over x_2}  \ \
|{\overline M}|^2_{ab \rightarrow cd}
\label{eq:correl-rate}
\end{eqnarray}
where $p_{_T{_i}}$ and $y_i$, $i = 3,4$, are respectively  the
transverse momenta and rapidities of the final state particles. The
momentum $k_{_T{_i}}$, $i = 3,4$, is the transverse momentum of the
parton $c$ (respectively, $d$) which emits particle $C$ (respectively,
$D$) of momentum $p_{_T{_i}}$, carrying a fraction  $z_i =
p_{_T{_i}}/k_{_T{_i}}$  of the parton momentum. The fragmentation
functions $D_{C/c}$ and $D_{D/d}$ depend on the collinear
factorization scale $M_{_F}$. For the production of hadrons we will
use the leading order functions of \cite{kkp}. When a photon is
detected in the final state it can be either produced  directly, in
which case the fragmentation function $D_{\gamma/c}(z,M_{_F})$ reduces
to a Dirac function $\delta(1-z)$ or it can be produced via
bremsstrahlung of a final state quark or gluon (see
Figure~\ref{fig:1f2f} for illustration). In the latter case we use the
BFG parametrization\footnote{These parametrizations are given at the
NLO of QCD.  Nevertheless we use them for our leading-logarithmic
studies for lack of recent leading-logarithmic parametrizations.}
of~\cite{bourhis}.  The structure  functions $F_{{a/A}}$ and
$F_{{b/B}}$ of the projectile and target, $A$ and $B$, depend on the
factorization scale $M$ and they are normalized to one nucleon. Our
standard  in the following study is the parametrization of
CTEQ6L~\cite{cteq6}.

The quantity  $|{\overline M}|^2_{ab \rightarrow cd}$ is the matrix
element squared, averaged over spin and color, of the partonic
sub-process $ab \rightarrow cd$. It depends implicitely on the
renormalisation scale $\mu$ through the strong coupling
$\alpha_s(\mu)$. In the following, unless otherwise specified, all
scales, $\mu$, $M$, and $M_{_F}$, are set equal to
($p_{_{T_3}}$ + $p_{_{T_4}}$)/2. Higher order corrections to
Eq.~(\ref{eq:correl-rate}) have been calculated and extensive
phenomenological studies have been made in the case of proton-proton
collisions~\cite{cfg1996,bgpw2001,diphox-program,wa70-gam-gam,e706-gam-gam,d0-gam-gam}.
They are briefly discussed in Section~\ref{se:nlo}.

\subsection{Initial state nuclear effects}
\label{se:nuc-eff}

The structure function of a nucleon in a nucleus is modified by
shadowing and antishadowing effects. These effects are hard to
calculate theoretically and in this study we will simply use a
parametrization of the parton distribution functions measured in
deep-inelastic scattering experiments of leptons off nuclei~\cite{nmc}
and Drell-Yan production in proton-nucleus reactions~\cite{e772}.  We
follow the approach of Eskola {\it et al.}~\cite{Eskola:1998df} who
tabulate a function $S_{a/{\rm A}}(x,M)$ which relates the parton
distributions in a nucleon $N$ to those in a nucleus $A$ via \bea
F_{a/{\rm A}}(x,M) = S_{a/{\rm A}}(x,M)\,F_{a/N}(x,M)
\label{eq:nucl-struc}
\ena   Unlike older parametrizations, it is to be noted that the
parametrization of~\cite{Eskola:1998df} treats quarks and gluons
separately and, furthermore, the factorization scale dependence is
taken into account. In this parametrization, shadowing effects reduce
the quark (by up to 10$\%$) and the gluon (by up to 30$\%$)
distributions in the nucleon below $x\lesssim 0.03$, while
anti-shadowing enhances the distributions by up to 10$\%$
between\footnote{The  precise $x$ values depend on the factorization
scale.}  $x = 0.03$ and $x = 0.4$.  Since observables usually involve
an integration over a rather large range in $x$, such
shadowing/antishadowing effects will not affect the predictions
very much. A summary of recent shadowing studies at LHC can be found
in~\cite{eskolaYR}.

Plugging Eq.~(\ref{eq:nucl-struc}) in Eq.~(\ref{eq:correl-rate}), with
proper account of isospin effects, one obtains the cross section
normalized per one nucleon in a nucleus. To obtain the counting rate
in an actual heavy ion experiment one needs to account for the number
of nucleon-nucleon scatterings occuring in a nucleus-nucleus
collision. This is done in a standard way using Glauber theory.  A
hard cross section for an $A A$ collision, with a given centrality
class $\cal C$ (equivalently impact parameter range), is obtained from
the corresponding hard cross section  via the ``binary scaling"
relation \bea {\sigma}^{hard}_{AA}|_{\cal C} = \langle N_{coll} \rangle\!|_{\cal C} \
{\sigma^{geo}_{AA} \over \sigma_{NN}} \ \sigma^{hard}_{NN}
\label{eq:nucl-sec}
\ena  where $\langle N_{coll}\rangle\!|_{\cal C}$ is the number of collisions at the
chosen centrality, $\sigma^{geo}_{AA}$ is the geometric cross section
obtained via the Glauber multiple scattering model and $\sigma_{NN}$
is the nucleon-nucleon cross section.  Details, as well as the
numerical values of the various terms, are given
in~\cite{photonYR}. In particular, for collisions with a centrality
less than 5$\%$ the estimate is $\langle N_{coll}\rangle\!|_{\cal C} = 1876$ and
$\sigma^{geo}_{AA} = 7745$~mb for lead-lead collisions at 5.5 TeV with
$\sigma_{NN} = 72$~mb. In the following we implicitely consider only
central collisions, with ${\cal C} \le 5\%$, for lead-lead collisions.
When quoting numbers of events for a given observable we assume the
standard luminosity for lead-lead collisions in ALICE, 
${\cal L} = 5.10^{26}$~cm$^{-2}$~sec$^{-1}$, 
and make the hypothesis that LHC is
running 30 days per year in the heavy ion mode.


\subsection{Medium-modified fragmentation functions}\label{subse:modelFF}

Much progress has been done over the last decade to better understand
the gluon radiation by hard partons travelling through dense QCD
media~\cite{wloss}. More recently, the important connection between
the medium-induced gluon spectra $dI/d\omega$ and the probability
distribution ${\cal P}$ in the energy loss has been made
explicit~\cite{bdms2001} and computed numerically soon
after~\cite{arleojhep,swprd}. However, it remains unclear how to
relate the parton energy loss mechanism to observable quantities.

For sufficiently large $\kt$ parton production, nevertheless, a clear
separation is achieved between the hard production process, with a
time scale $\cO{\kt^{-1}}$, the effects of the medium, $\cO{t_{\rm
med}}$, and the fragmentation mechanism, $\cO{\kt/\Lambda^2}$,
\begin{equation}\label{eq:timescales}
\frac{1}{\kt} \ll t_{\rm med} \ll \frac {\kt}{\Lambda^2}
\end{equation}

Provided the hierarchy~(\ref{eq:timescales}) is justified, it is
sensible to model the energy loss effects at the level of
fragmentation functions. In the present study, we shall follow the
model suggested in  Ref.~\cite{Wang:1996} in which the energy $\eps$
lost by the hard parton leads to a rescaling of the momentum fraction
$z_d$
\begin{equation}\label{eq:shift}
z_d = \frac{p_{_{Td}}}{k_{_{Td}}} \qquad \to \qquad z_d^* =
\frac{p_{_{Td}}}{k_{_{Td}} - \epsilon} = \frac{z_d}{1 -
\epsilon/k_{_{Td}}}
\end{equation}
in presence of a QCD medium. Consequently, the medium-modified
fragmentation functions  $D_{D/d}^{\rm med}(z_d, M_{_F}, k_{_{Td}})$
may simply be expressed as a function of the standard (vacuum)
fragmentation functions $D_{D/d}(z_d, Q^2)$,
\begin{equation}
\label{eq:modelFF}
z_d\,D_{D/d}^{\rm med}(z_d, M_{_F}, k_{_{Td}}) = \int_0^{k_{_{Td}} (1
  - z_d)} \, d\epsilon \,\,{\cal P}_d(\epsilon, k_{_{Td}})\,\,\,
z_d^*\,D_{D/d}(z_d^*, M_{_F}).
\end{equation}
Here, ${\cal P}_d(\epsilon, k_{_{Td}})$ denotes the probability for
the parton with energy  $k_{_{Td}}$ to lose the energy
$\epsilon$~\cite{bdms2001}, which has been given a simple analytic
parametrization in~\cite{arleojhep} which we shall use in the present
calculations. The calculation of Ref.~\cite{arleojhep} is based on the
medium-induced gluon spectrum determined by Baier, Dokshitzer,
Mueller, Peign\'e and Schiff (BDMPS) including $\cO{1/k_{_{Td}}}$
corrections~\cite{Baier:1996kr,Baier:1998kq}. Such a model was shown
to describe successfully hadron production in semi-inclusive DIS
reactions off nuclear targets~\cite{arleodis}.  The BDMPS framework
should be particularly suited when the number of scatterings incurred
by the hard parton in the QCD medium (opacity) is large.  While thick
and dense media are indeed expected to be produced in nuclear
collisions at LHC energy, we note however that such a calculation may
not properly describe the energy loss process for partons produced
close to the surface. Let us mention that the probability distribution
at {\it finite} opacity was also determined in the soft limit
($k_{_{Td}}\to\infty$) in Ref.~\cite{swprd} which lead to recent
phenomenological applications at RHIC and
LHC~\cite{Eskola:2004cr,Dainese:2004te}.

The BDMPS energy loss distribution is characterized by the energy
scale~\cite{Baier:1996kr}
\begin{equation}\label{eq:omc}
\omega_c = \frac{1}{2}\,\hat{q}\, L^2
\end{equation}
where the so-called gluon transport coefficient $\hat{q}$ reflects the
medium gluon density~\cite{Baier:1996sk} and $L$ the  length of matter
covered by the hard parton in the medium. Note that $\hat{q}$ in
(\ref{eq:omc}) has to be seen as a time averaged quantity $\langle
\hat{q} \rangle$ to take properly into account the longitudinal
expansion of the produced medium~\cite{swprl}.

For the calculations to come, a qualitative estimate of $\omega_c$ for
the dense medium produced in lead-lead collisions at LHC energy is
needed. The transport coefficient~$\hat{q}$ is directly related to the
gluon density whose increase from RHIC to LHC is of order $6-7$ in
hydrodynamical calculations~\cite{Eskola:2004cr,priv_com}, while a
smaller increase $2-4$ in the hadron multiplicity at mid-rapidity
(also linked to $\hat{q}$) is predicted by several models (a review
can be found in~\cite{Armesto:2000xh}). Using the estimate based on
the pion $p_{_T}$ spectra measurements at RHIC, ${\omega_c}_{|_{_{\rm
RHIC}}} \simeq 10 - 20$~GeV, within the same
framework~\cite{arleojhep,Arleo:2004ri} we shall take throughout this
study the rather conservative choice\footnote{This choice is motivated
by the fact that the gluon distribution at very small $x$ could evolve
more slowly than seen so far at HERA due to possible gluon
saturation. Moreover, we want our predictions to be seen as lower
estimates as far as medium effects are concerned.}, $\omega_c =
50$~GeV.

Let us remind the reader that the goal here is not to provide
quantitative predictions but rather to show typical trends one could
expect in \gamgam and \gampi correlations in heavy-ion collisions at
the LHC. Therefore, our conclusions should not depend much on the
precise value we assume for the energy loss parameter $\omega_c$.

\begin{figure}[!ht]
\begin{center}
\includegraphics[height=14.0cm]{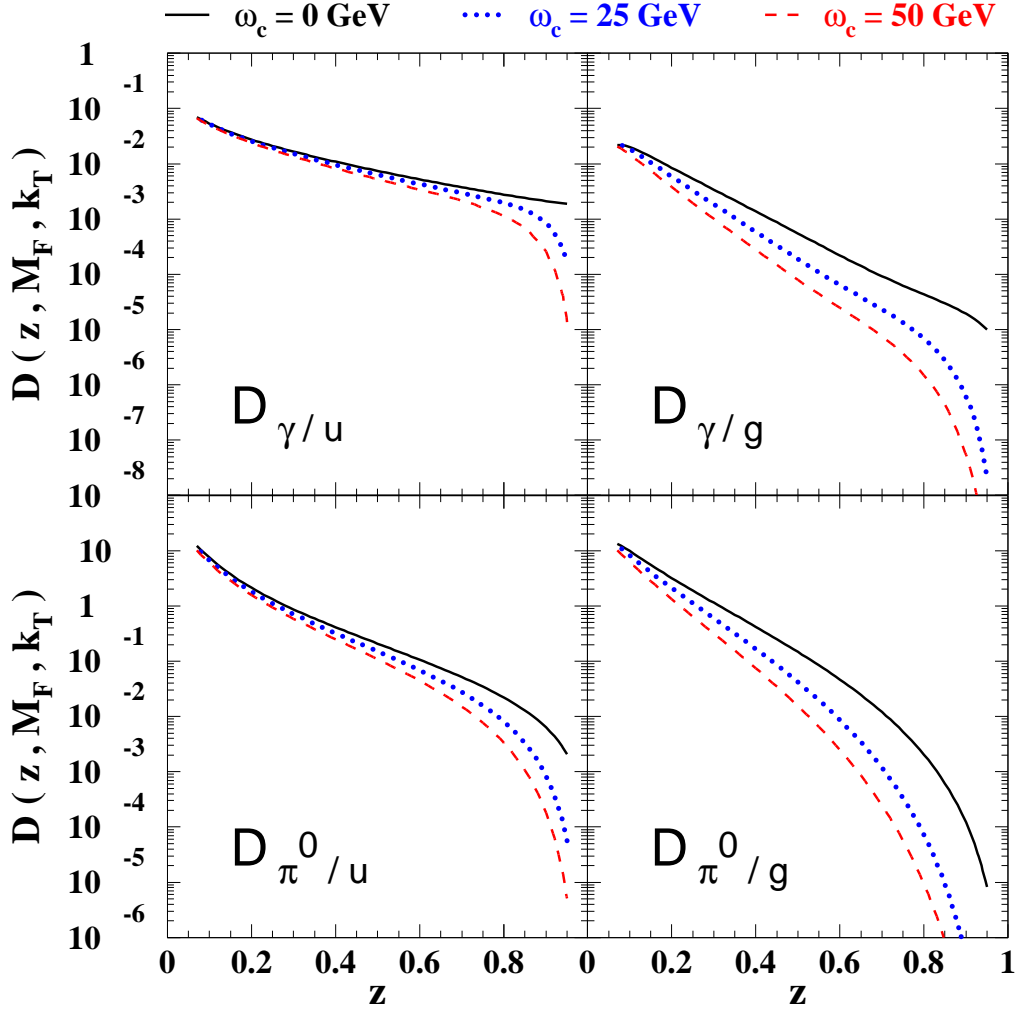}
\end{center}
\vspace{-0.8cm}
\caption{Medium-modified fragmentation functions $D_{D/d}^{\rm med}
(z_d, M_{_F}, k_{_{Td}})$ for various energy loss scales, $\omega_c =
0$~(vacuum), 25 and 50~GeV. The parton energy is $\kt = 50$~GeV and
the fragmentation scale is set to $M_{_F} = p_{_T} /2$.  }
\label{fig:quarktophoton}
\end{figure}

Since fragmentation functions fall steeply with $z$, even a small
shift  $\Delta z_d=z_d^*-z_d\approx~z_d\,\epsilon/k_{_{Td}}$ in
Eq.~(\ref{eq:shift}) may substantially affect the fragmentation
process due to parton energy loss.  This can be seen for  instance in
Figure~\ref{fig:quarktophoton} where the fragmentation
functions into a photon and into a neutral pion, using
respectively the BFG~\cite{bourhis} and KKP LO~\cite{kkp}
parametrizations, are computed for $k_{_T} = 50$~GeV up quark and
gluon traversing the medium ($\omega_c = 25, 50$~GeV) or not
($\omega_c = 0$~GeV).

First, Figure~\ref{fig:quarktophoton} indicates that medium effects
prove stronger for gluon than for quark fragmentation. The origin is
actually twofold. First, hard gluons lose more energy than quarks do
from their larger color charge ($C_g = 3$, $C_q = 4/3$). Moreover, the
quenching of medium-modified fragmentation
functions~Eq.~(\ref{eq:modelFF}) increases with the slope of vacuum
fragmentation functions, much steeper in the gluon channel. Finally,
we observe that the effects of parton  energy loss become more
pronounced as $z$ gets larger, due to the restricted available phase space in
Eq.~(\ref{eq:modelFF}).

\begin{figure}[!ht]
\begin{center}
\includegraphics[height=14.0cm]{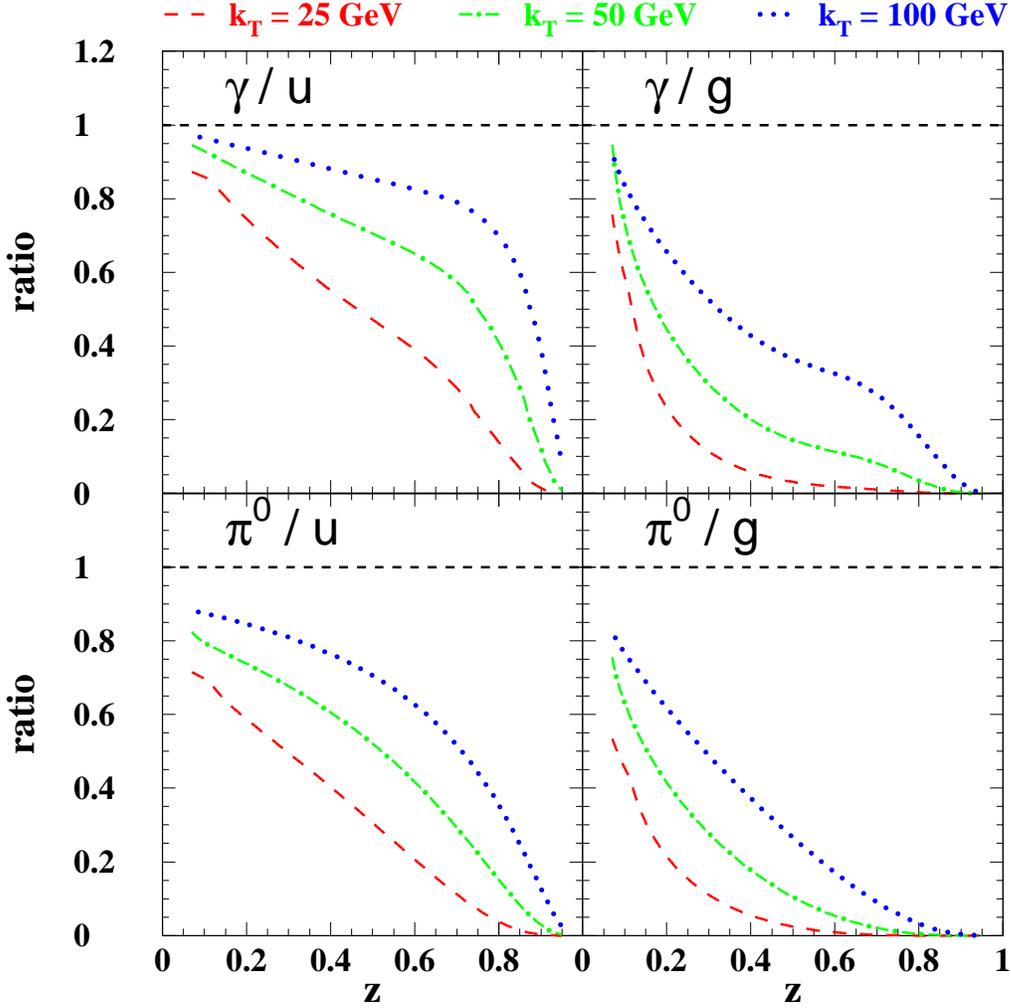}
\end{center}
\vspace{-0.8cm}
\caption{   Ratio of medium-modified ($\omega_c = 50$~GeV) over vacuum
($\omega_c = 0$~GeV) fragmentation functions for various parton
energy, $\kt = 25, 50$~and 100~GeV. The fragmentation scale is set to
$M_{_F} = p_{_T} /2$.  }
\label{fig:quarktophoton2}
\end{figure}  

The medium-modified fragmentation functions depend now explicitely on
the parton energy, $\kt$. To show the sensitivity of the medium
effects on the parton energy, the ratio of medium (using $\omega_c =
50$~GeV) over vacuum fragmentation functions is determined for $k_{_T}
= 25, 50$ and 100~GeV quarks and gluons. As can be seen in
Figure~\ref{fig:quarktophoton2}, medium effects will be magnified as
the parton energy is getting smaller. When $\kt$ becomes too small as
compared to $\omega_c$, however, the picture of a hard parton
penetrating the soft medium is no longer correct and the applicability
of the energy loss framework becomes doubtful. It was shown for
instance in~\cite{arleojhep} that the eikonal approximation is
explicitely broken down for parton energy smaller than half the scale
$\omega_c$. This has to be seen as the lower limit for the most
energetic photon transverse momentum. On the other hand, in the high
energy limit $\kt\gg\omega_c$ and thus $z^*\simeq z$, the medium
effects vanish and the ratio approaches one.

\section{The correlations}\label{se:correlations}

\subsection{Observables}

From the kinematical variables available in Eq.~(\ref{eq:correl-rate})
we can construct the following observables:\\  -- the invariant mass
of the particle pair,  \bea m^2_{34}&=& 2\ (p_{_T{_3}} p_{_T{_4}} {\rm
ch}(y_3-y_4)  -  \vec p_{_T{_3}} \cdot \vec p_{_T{_4}}) \nonumber \\
&=& 2 \ z_3\ z_4\ k^2_{_T} \ ({\rm ch}(y_3-y_4) + 1)
\label{eq:invmass}
\ena  -- the transverse momentum of the pair,  \bea {q_{_T}} &=&
|{\vec p_{_T{_3}}} + {\vec p_{_T{_4}}}| \nonumber \\ &=& k_{_T}\ |z_3
-z_4|
\label{eq:pairpt}
\ena  -- the relative transverse momentum of the particles (also
called momentum balance~\cite{cfg1996,abdfs1984})  \bea z_{34} &=& -
{\vec p_{_T{_3}} \ .\ \vec p_{_T{_4}} \over p^2_{_T{_3}} } \nonumber
\\ &=& {z_4 \over z_3}
\label{eq:scaledp34}
\ena  where $k_{_T}$ is the common value of the transverse momentum of
the final state partons. For completeness we quote here the
expressions of the $x_i$ values of the initial partons,  \bea x_1 &=&
{1\over \sqrt{s}}\ \left({p_{_T{_3}}\over z_3} e^{y_3} +{\pt4\over
z_4} e^{y_4} \right) \nonumber \\ x_2 &=& {1\over \sqrt{s}}\
\left({p_{_T{_3}}\over z_3} e^{-y_3} +{\pt4\over z_4}  e^{-y_4} \right)
\label{eq:x1x2}
\ena

In the case of \gampi correlations, one has $z_3 = 1$ when the photon
is produced directly. Fixing furthermore the rapidity of the photon
and the pion in a narrow range around 0, for example, we are left with
two independent kinematical variables $z_4$ and $\kt$ and the
expressions of the observables defined above considerably
simplify. One has:  \bea m^2_{34} &=& 4\ k^2_{_T}\ z_4 \nonumber \\
q_{_T} &=& k_{_T} \ |1-z_4| \nonumber \\ z_{34} &=& z_4
\label{eq:simple}
\ena which show a straightforward relation between the fragmentation
variable $z_4$ and the observables. 

When studying the observables we integrate over $\kt$ above
a given value. Since the cross section is rapidly falling when $\kt$
is increasing the effective transverse momentum will remain close to
its minimum value leaving $z_4$ as the only effective variable. From
the behavior of the above observables one should get constraints on
the behavior of the fragmentation function if we assume that the
structure functions are precisely known.

When the photon is produced via bremsstrahlung or in the case of \pipi
correlations the above simple situation is somewhat smeared because
$z_3$ is now a relevant kinematical variable. However, when studying
asymmetric configurations with a large $p_{_T{_3}}$ particle on one
side and a small $\pt4$ particle on the other side, trigger bias
effects will force a large value of $z_3$ and some correlation is
still expected between the observables of Eqs.~(\ref{eq:invmass}) to
(\ref{eq:scaledp34}) and the $z_4$ dependent fragmentation function.

On the other hand, considering \gamgam correlations, when both photons
are produced directly, one has an over constrained system  with $z_3=
z_4 =1$ and the distributions in $z_{34}$ and in $q_{_T}$ reduce to
Dirac $\delta$ functions while the invariant mass $m^2_{34} = 2
k^2_{_T} ({\rm ch}(y_3-y_4) + 1)$  is regular.  It is obvious that
higher order corrections will smear the $\delta$-function
singularity. More precisely, $z_{34} = 1$, or equivalently $q_{_T} =
0$, is an infrared sensitive point and an accurate prediction of the
behavior of these observables near this point will require the
resummation of large $\ln^2(q^2_{_T}/s)$  and $\ln(q^2_{_T}/s)$ terms.

\subsection{Kinematical cuts}

We study a basic perturbative QCD (pQCD) mechanism modified by the
presence of a dense environment. It is necessary to insure that the
particles we observe are decay fragments of jets and are not produced
by secondary collisions. Recent studies~\cite{photonYR} in the
framework of perturbative QCD for primary collisions and  a
hydrodynamic model to describe secondary collisions have shown that,
at LHC, particles produced above $p_{_T} = 5$~GeV are of pQCD
origin. We therefore impose a minimum transverse momentum of 5~GeV on
the particles from which we construct the various correlation
observables.

To study a large domain in the fragmentation variable $z$ it is
necessary to consider asymmetric configurations. Another constraint is
to be able to distinguish  photons from pions which requires, for
ALICE for example, \ptgamma $>$ 25~GeV. On the other hand, to have a
reasonable counting rate, one should not go to too high values of
\ptgamma.  Besides, if jets are too energetic, energy loss effects
will be small and difficult to observe.

In the subsequent studies, apart from various $p_{_T}$ distributions,
we will look at the three distributions in Eqs.~(\ref{eq:invmass}) to
(\ref{eq:scaledp34})  with \ptgamma $>$ 25~GeV and \ptpi $>$ 5~GeV to
satisfy the above criteria. We will also consider higher cuts in
\ptgamma  to probe the sensitivity of the energy loss mechanism on the
jet energy.  When displaying the distributions we always assume
photons and  pions are produced in an interval of $\delta y = \pm\
0.5$ unit of rapidity around  $y=0$.

\section{Phenomenology of \gampi correlations}\label{se:gampi}

\subsection{Dynamical components}

Before studying the shape of correlation functions in lead-lead
collisions it is worthwhile considering into some details the case of
proton-proton scattering to better understand the dynamics of the
reaction. As above mentioned (see Figure~\ref{fig:1f2f}), the photon
can be produced directly and only the recoiling jet fragments into a
pion (labeled 1f, open squares in the following figures), or  both the
photon and the pion are produced by fragmentation of  partons (labeled
2f, full squares).

\begin{figure}[!ht]
\begin{center}
\includegraphics[height=14.0cm]{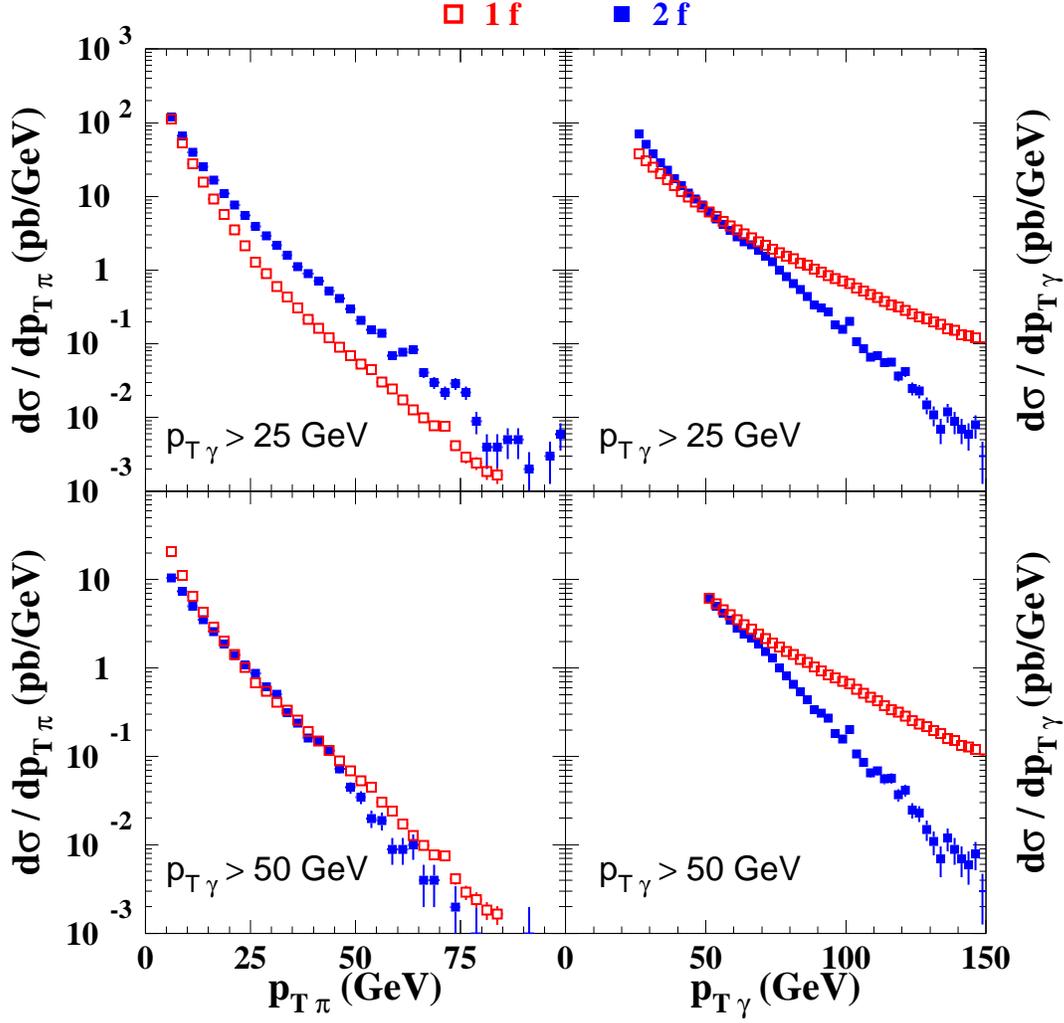}
\end{center}
\vspace{-0.8cm}
\caption{   The 1f and 2f components for various kinematical
configurations in \gampi production for proton-proton scattering at
$\sqrt{s} = 5.5$~TeV. {\it Top:} The \ptgamma and the \ptpi
distributions for the cuts \ptgamma $>$ 25~GeV and \ptpi $>$
5~GeV. {\it  Bottom:} Same as above for \ptgamma $>$ 50~GeV and  \ptpi
$>$ 5~GeV.  }
\label{fig:bkgd-pp}
\end{figure}  

The relative weight of each mechanism depends crucially on the
transverse momentum of the produced particles.  This is illustrated in
Figure~\ref{fig:bkgd-pp}. On the right panels one displays the
production rate as a function of \ptgamma: typically, the 2f process
dominates when \ptgamma $<$ 50 GeV but decreases much faster than the
1f component as \ptgamma increases and becomes negligible for photon
transverse momenta above 100~GeV. Therefore, changing the \ptgamma cut
will affect dramatically the 1f and 2f relative contributions to
\gampi production.

As shown in Figure~\ref{fig:bkgd-pp}  ({\it left}), the pion \ptpi
spectrum is dominated by the 2f process in a  large \ptpi range when
the photon energy is not too large (\ptgamma$\ge$ 25~GeV) while
applying a higher cut (\ptgamma$\ge$ 50~GeV) in \ptgamma results into
a comparable magnitude for both mechanisms on a wide \ptpi
~domain. Unlike the \ptgamma distribution the relative weight of the
2f process is increasing with the transverse momentum of the pion in
the kinematical domain shown. Indeed, the 1f component is disfavored
at large \ptpi as it requires more energetic photons (\ptgamma $\ge$
\ptpi), while the 2f process allows the photon to keep a small
transverse momentum, slightly above the 25~GeV or 50~GeV cut.

Experimentally, these two contributions may be disentangled by means
of calorimetry techniques using appropriate isolation
criteria. However, the large multiplicity reached in high energy
heavy-ion collisions prevent one from using such techniques. It should
also be reminded that when performing a full NLO study, the
distinction between the leading-order fragmentation and the
next-to-leading order direct component is arbitrary and depends on the
fragmentation scale $M_F$, only the sum of these components being
meaningful and ideally scale independent~\cite{bgpw2001}.

Medium effects may change considerably whether one or the other
process  dominates since the typical parton energy, $\kt =
p_{_{T_\gamma}}/z_{_3}$, is quite different for 1f ($z_{_3} = 1$) and
2f ($z_{_3} < 1$).  Naively, the effects of parton energy loss should
be stronger when {\it both} the pion and the photon come from the hard
parton fragmentation. However, this is not necessarily true since the
parton energy is actually much greater in the 2f channel, for which
medium effects prove weaker (see end of
Section~\ref{subse:modelFF}). We shall come back to these observations
when discussing the proton-proton and lead-lead spectra in the next
section.

\subsection{Distributions}

\begin{figure}[!ht]
\begin{center}
\includegraphics[width=16cm]{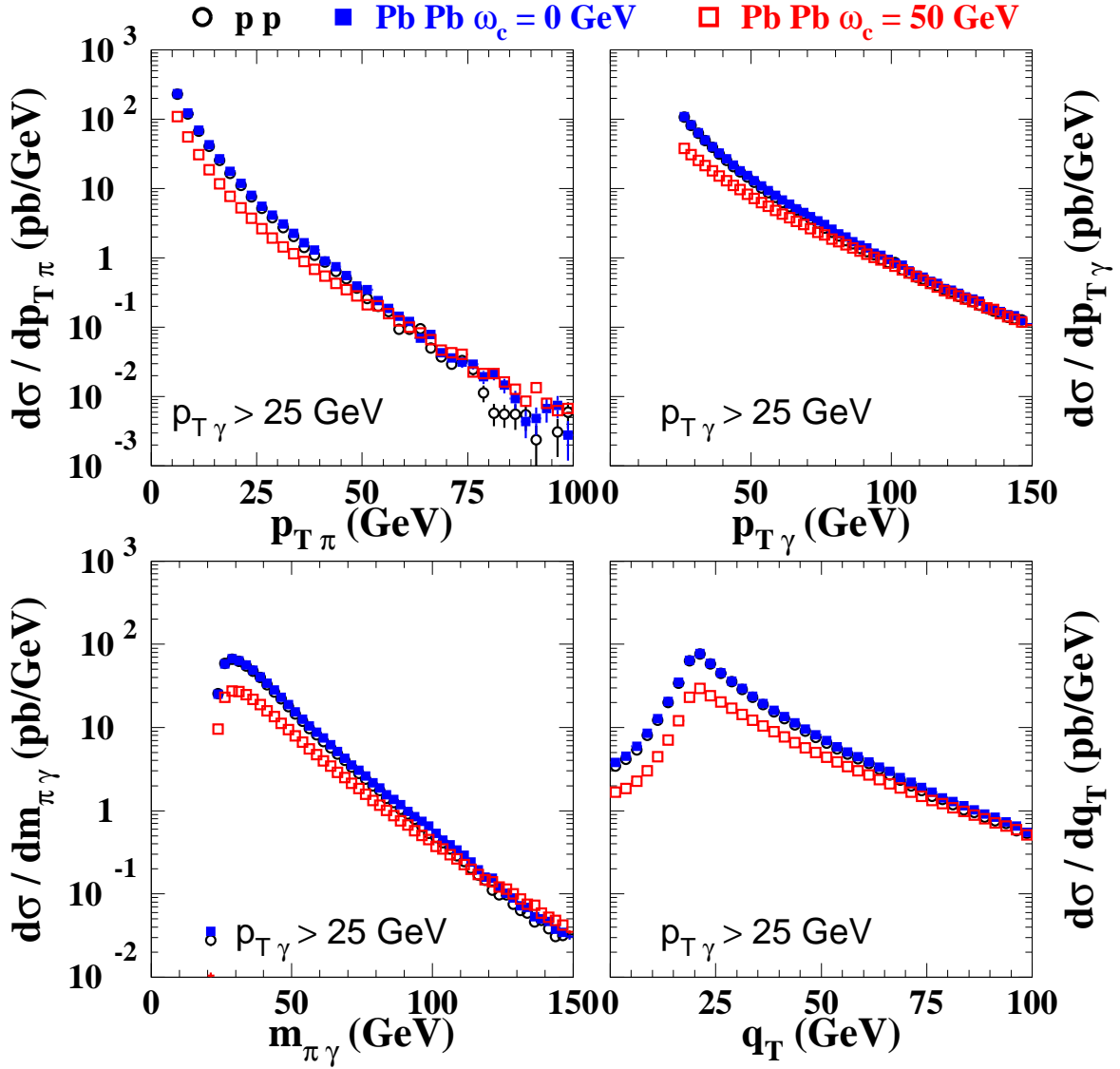}
\end{center}
\vspace{-0.8cm}
\caption{   The four distributions in \gampi production defined in the
text  for proton-proton (open dots) and lead-lead scattering (no
energy loss: black squares; with energy loss: open squares) at
$\sqrt{s} = 5.5$~TeV. Both the photon and the pion are produced at
rapidity [-0.5, 0.5] and the following cuts are imposed: \ptgamma $>$
25~GeV and \ptpi $>$ 5~GeV.  }
\label{fig:quarteron_pi0_gam_25_05}
\end{figure}

In Figure~\ref{fig:quarteron_pi0_gam_25_05} we discuss four
distributions, respectively in the pion transverse momentum \ptpi, the
photon transverse momentum \ptgamma, the \gampi invariant mass
$m_{\pi\gamma}$ and the transverse momentum of the pair $q_{_T}$. We
impose the following cuts: ~\ptpi$\ge 5$~GeV and \ptgamma$\ge
25$~GeV. In each case three curves are displayed: proton-proton
scattering (open dots), lead-lead scattering with shadowing but
without energy loss (full squares) and lead-lead scattering with
shadowing and energy loss using $\omega_c = 50$ GeV (open squares).

It is clear that shadowing and isospin effects do not modify the
distributions very much: a small antishadowing effect can be observed
at large transverse momenta or at high invariant mass due to the fact
that the kinematics then becomes sensitive to larger $x$ partons in
the nuclei. Energy loss effects are quite visible
particularly at the low $p_{_T}$ values of the pion or the photon. 
On the other hand, to
produce a pion at high transverse momentum requires a parton
with large $\kt$  for which the energy loss is expected to be smaller.
We observe, accordingly, that the spectrum in lead-lead collision
tends to approach the proton-proton spectrum as \ptpi increases.  The
medium effects are also particularly visible on the spectrum as a
function \ptgamma: as long as the photon is produced directly (1f),
the \ptgamma spectrum reflects the energy of the parton, $\kt = $
\ptgamma, which eventually fragments into the pion. Again, the
quenching will be maximal for small \ptgamma (small $\kt$) while at
asymptotic energies, parton energy loss will have no observable
consequence. Similar behavior is observed in the invariant mass
distribution: small masses correspond to low $\kt$ partons and
therefore lead to a stronger suppression. One may notice, in passing,
the rather large counting rates: with the numbers given in
Section~\ref{se:nuc-eff} we estimate that about 1300  \gampi pairs with
a 100~GeV invariant mass will be produced, per year, in ALICE.

Perhaps, more interesting is the $q_{_T}$ spectrum which exhibits a
maximum when the pion and the photon transverse momenta lie just above
the imposed kinematic threshold, which is located at the difference
between the \ptgamma and the \ptpi cut, 20~GeV. Above 20 GeV, the
distribution is reminiscent of the \ptgamma and the $m_{\pi\gamma}$
distribution and, in particular, the larger the $q_{_T}$ the weaker
the energy loss effect. Similarly, the energy loss effects will tend
to be smaller at very small $q_{_T}\ll$~20~GeV as the pion transverse
momentum and thus the 2f contribution --~less affected by the
medium~-- increases with decreasing $q_{_T}$. Therefore, we expect the
medium effects to be maximal for $q_{_T}$ roughly around the
difference of the transverse momentum cuts.

\begin{figure}[!ht]
\begin{center}
\includegraphics[width=16cm]{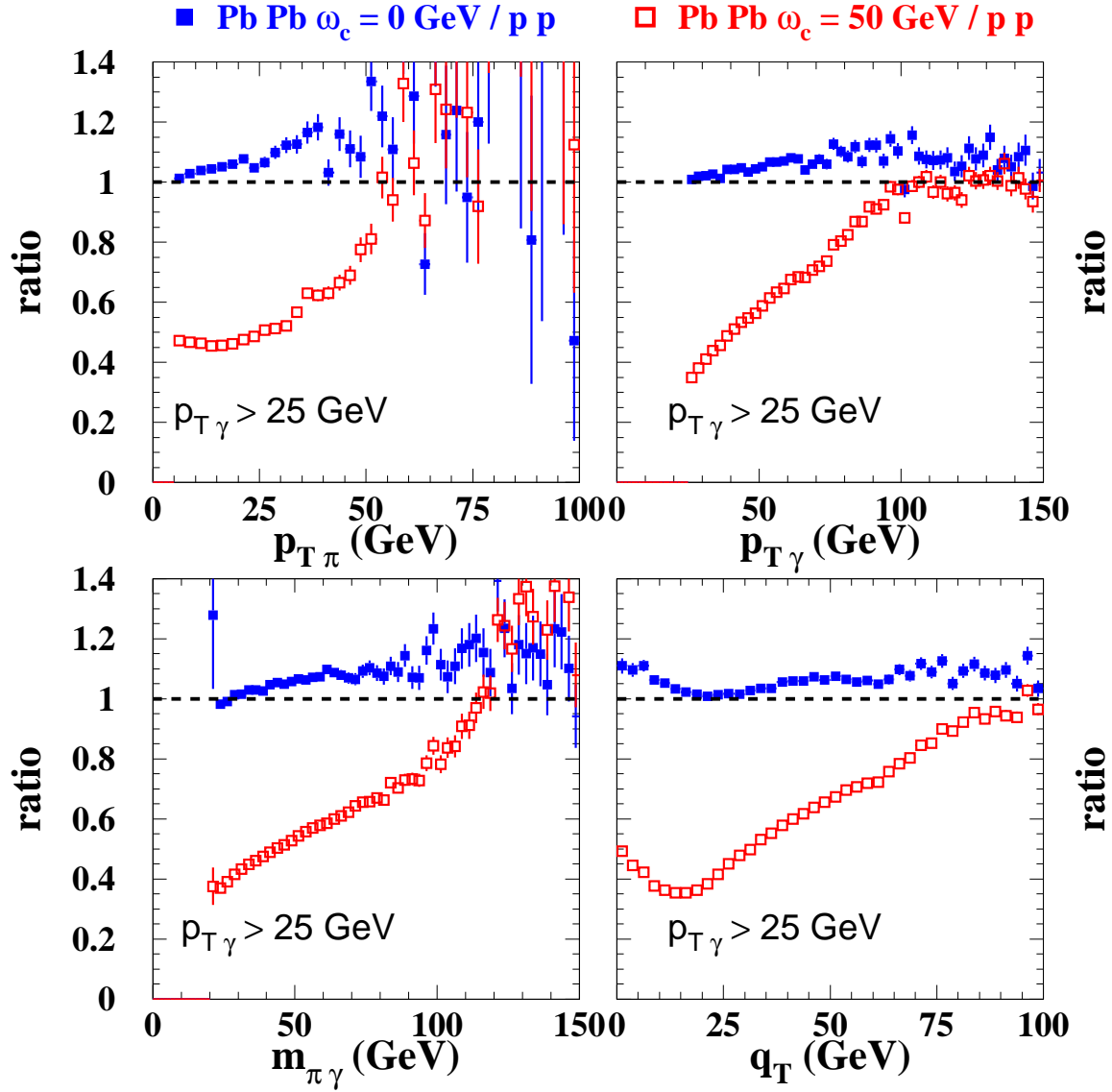}
\end{center}
\vspace{-0.8cm}
\caption{ Same as Figure~\ref{fig:quarteron_pi0_gam_25_05} but the
distributions are normalized to proton-proton scattering.  }
\label{fig:quarteron_pi0_gam_25_05_ratio}
\end{figure}

All these features are best seen when normalizing, to the
proton-proton distributions, the lead-lead distributions with energy
loss (open squares) or without  energy loss (full squares) in
Figure~\ref{fig:quarteron_pi0_gam_25_05_ratio}. In all cases it
appears that the observables are affected by antishadowing, and not
shadowing, but this remains a small effect, less than 10\% in general.
Energy loss effects, on the other hand, modify the distributions much
more drastically.  The \ptpi ~spectrum is suppressed by about 50\%
below \ptpi $=$ 25 GeV but no suppression occurs above 50 GeV. As for
the \ptgamma distribution, the suppression is maximum at low
transverse  but  is monotonously reduced as the momentum increases. In
the $q_{_T}$ spectrum the change of slope discussed above it is
particularly noticeable.

\begin{figure}[!ht]
\begin{center}
\includegraphics[width=16cm]{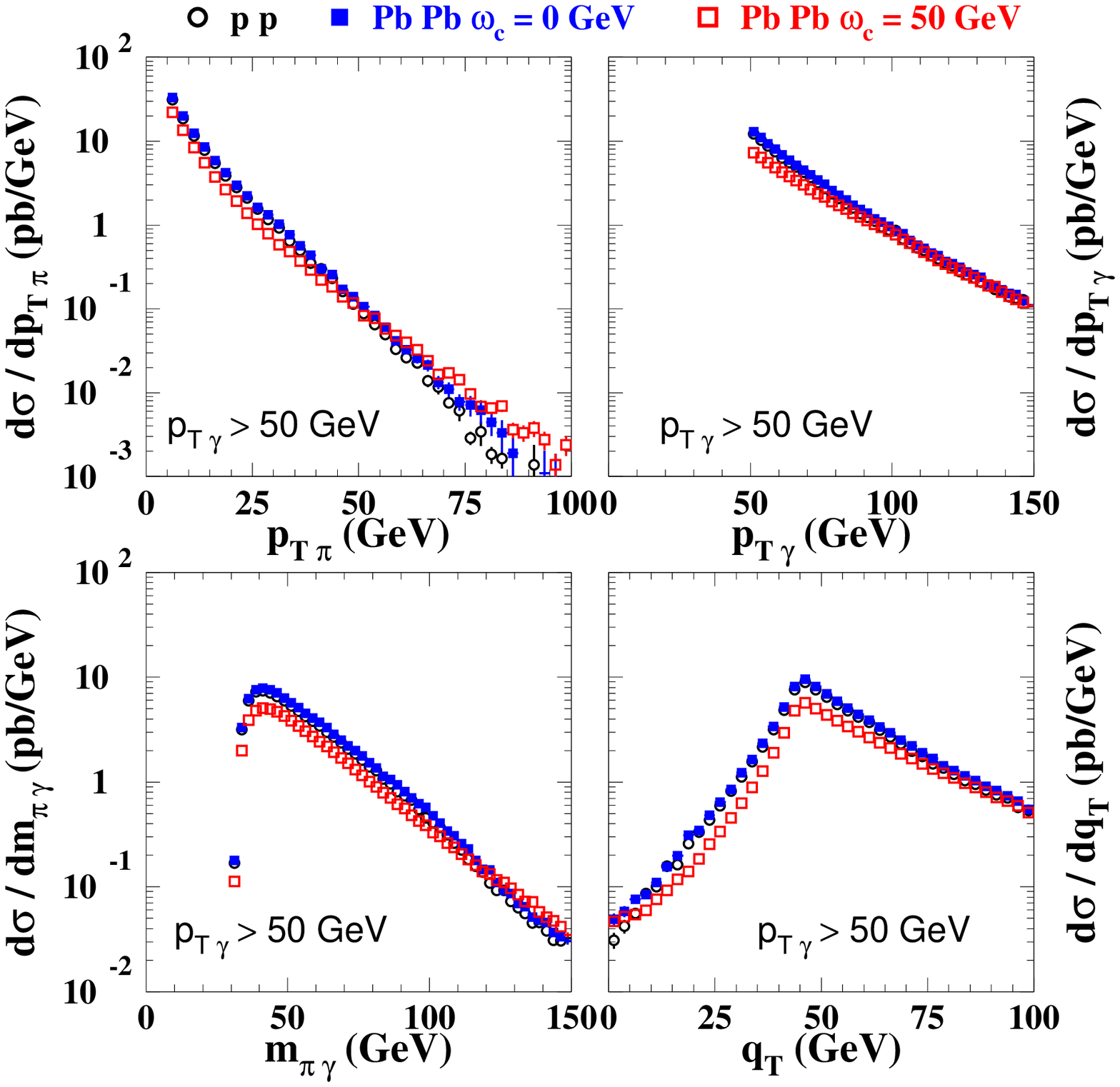}
\end{center}
\vspace{-0.8cm}
\caption{   Same as Figure~\ref{fig:quarteron_pi0_gam_25_05} with the
following cuts: ~\ptgamma $>$ 50~GeV and \ptpi $>$ 5~GeV.}
\label{fig:quarteron_pi0_gam_50_05}
\end{figure}

\begin{figure}[!ht]
\begin{center}
\includegraphics[width=16cm]{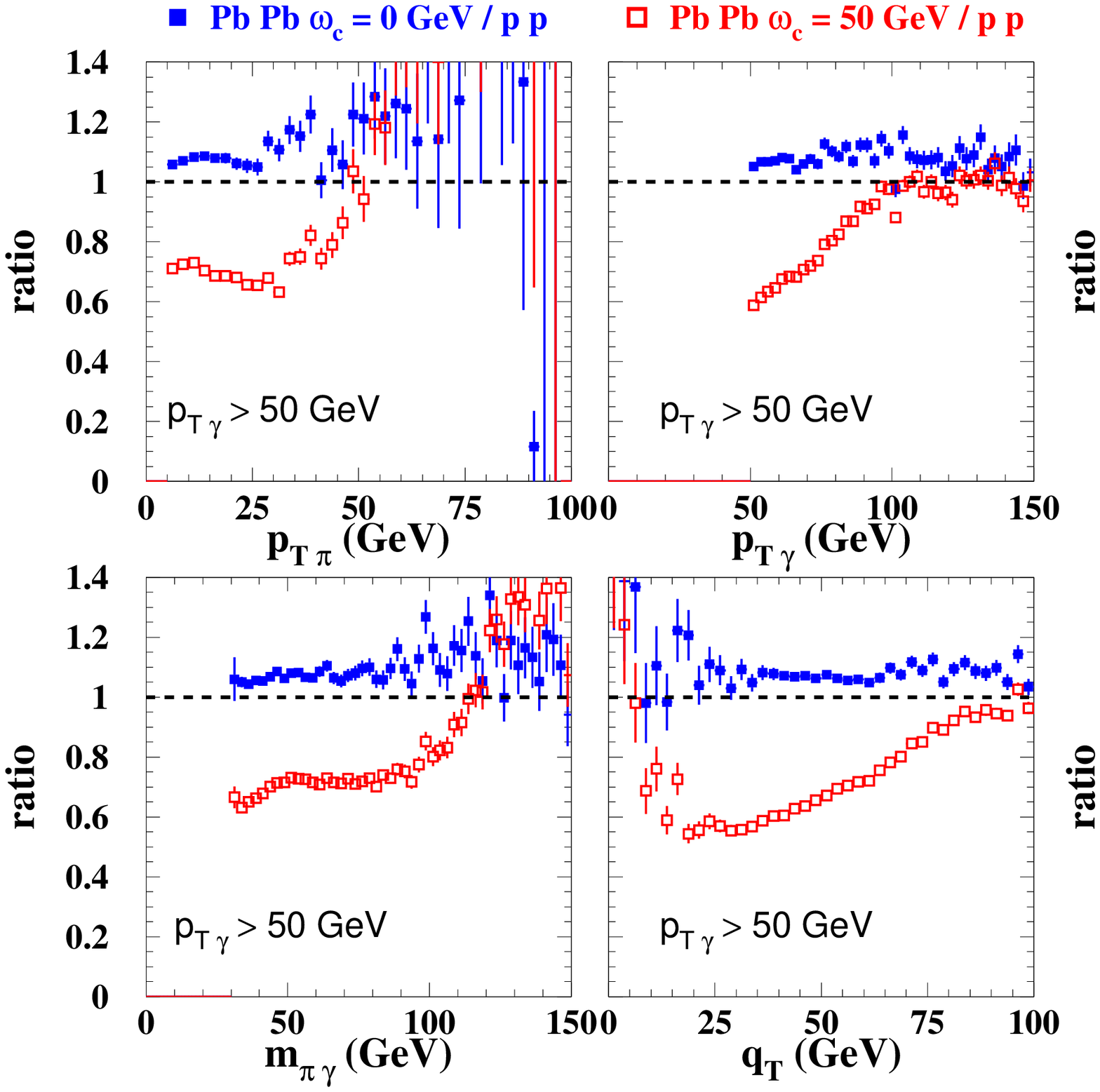}
\end{center}
\vspace{-0.8cm}
\caption{   Same as Figure~\ref{fig:quarteron_pi0_gam_25_05_ratio}
with the following cuts: ~\ptgamma $>$ 50~GeV and \ptpi $>$ 5~GeV.}
\label{fig:quarteron_pi0_gam_50_05_ratio}
\end{figure}

The same spectra are computed in
Figure~\ref{fig:quarteron_pi0_gam_50_05} assuming a larger cut for the
photon transverse momentum, \ptgamma$\ge 50$~GeV.  Although the 1f
contribution becomes relatively more important, one observes  similar
features as before. Again, the normalized $q_{_T}$ spectrum shows a
clear minimum in Figure~\ref{fig:quarteron_pi0_gam_50_05_ratio} around
$q_{_T} \simeq 30$~GeV, under which the 2f contribution starts to
dominate. The quenching of these spectra proves less pronounced --~the
ratio decreases down to 0.6 in the $q_{_T}$ spectra to be compared to
0.35 before~-- since the initial parton  energy is twice as
large. Finally, let us note that the counting rate drops  by a factor
$5$ to $10$ when increasing the photon cut from 25~GeV to 50~GeV.


\begin{figure}[!ht]
\begin{center}
\includegraphics[height=14.0cm]{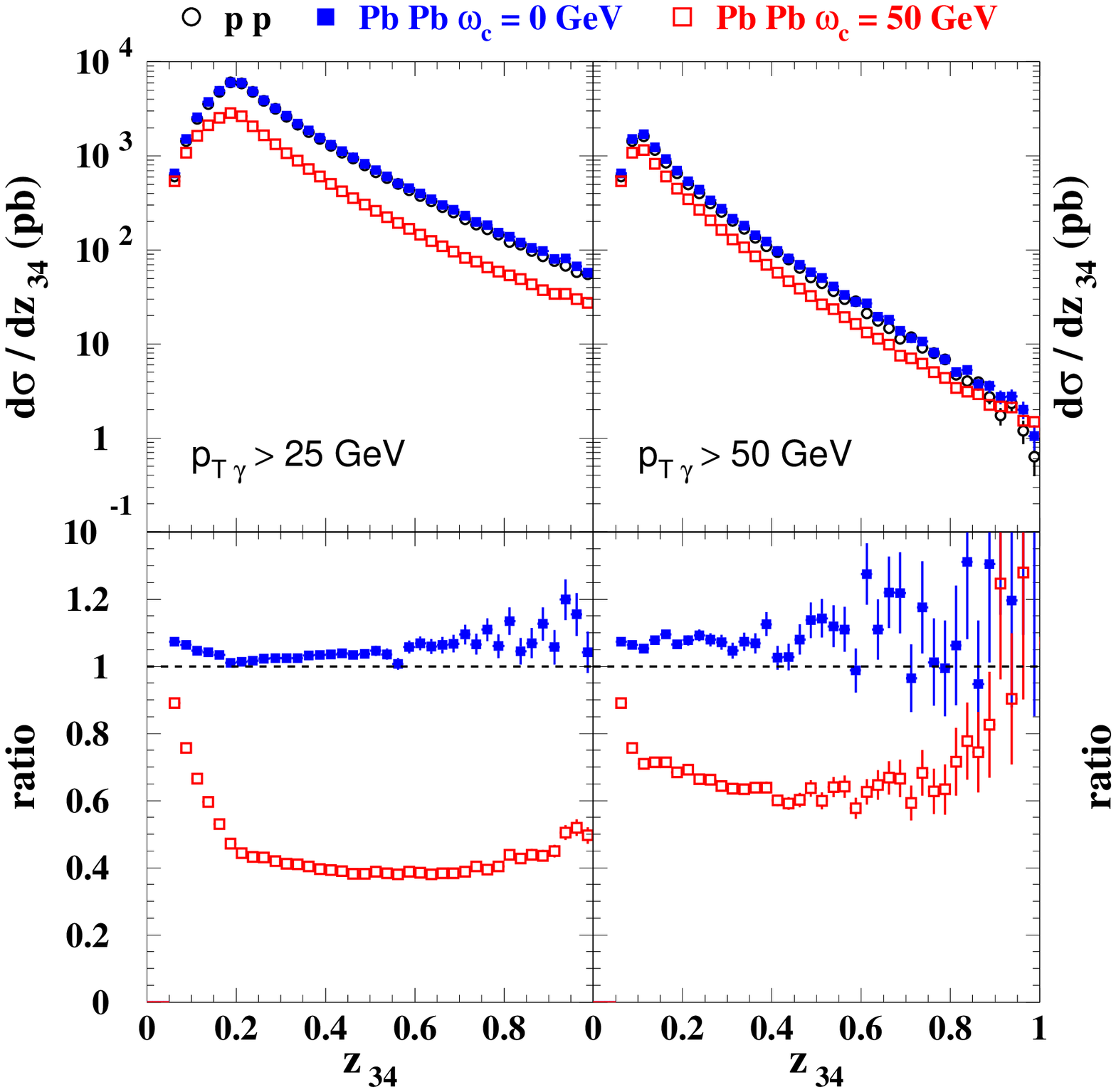}
\end{center}
\vspace{-0.8cm}
\caption{   The $z_{_{3 4}}$ distribution in \gampi production for
proton-proton (open dots) and lead-lead scattering (no energy loss:
black squares; with energy loss: open squares) at $\sqrt{s} =
5.5$~TeV. Both the photon and the pion are produced at rapidity [-0.5,
0.5] and the following cuts are imposed: \ptgamma $>$ 25~GeV and \ptpi
$>$ 5~GeV ({\it left}) and \ptgamma $>$ 50~GeV and \ptpi $>$ 5~GeV
({\it right}).  {\it Bottom:} The same distributions normalized to the
proton-proton case.  }
\label{fig:z}
\end{figure}

We turn now to a detailed discussion of the distribution in the
momentum balance $z_{_{3 4}}$ for both cuts \ptgamma$\ge 25$~GeV and
\ptgamma$\ge 50$~GeV, keeping \ptpi$\ge 5$~GeV as before
(Figure~\ref{fig:z}, {\it top}). The maximum of these distributions is
reached for $z_{_{3 4}} = 0.2$ and 0.1, i.e. the ratio of the pion
over the photon transverse momentum cuts. Smaller (larger) $z_{_{3
4}}$ values are obtained by increasing the photon (pion) transverse
momentum. Looking at the medium effects (Figure~\ref{fig:z}, {\it
bottom}), one observes rather structureless features: as compared to
the proton-proton case the spectrum is reduced to 40\% (respectively
60\%) over most of the $z_{_{3 4}}$ range when the cut \ptgamma $>$
25~GeV (respectively \ptgamma $ > 50$~GeV) is imposed.  Below $z_{_{3
4}} \le 0.2$, the suppression is not as strong since the photon
energy, hence $\kt$, is getting larger.

\begin{figure}[!ht]
\begin{center}
\includegraphics[height=14.0cm]{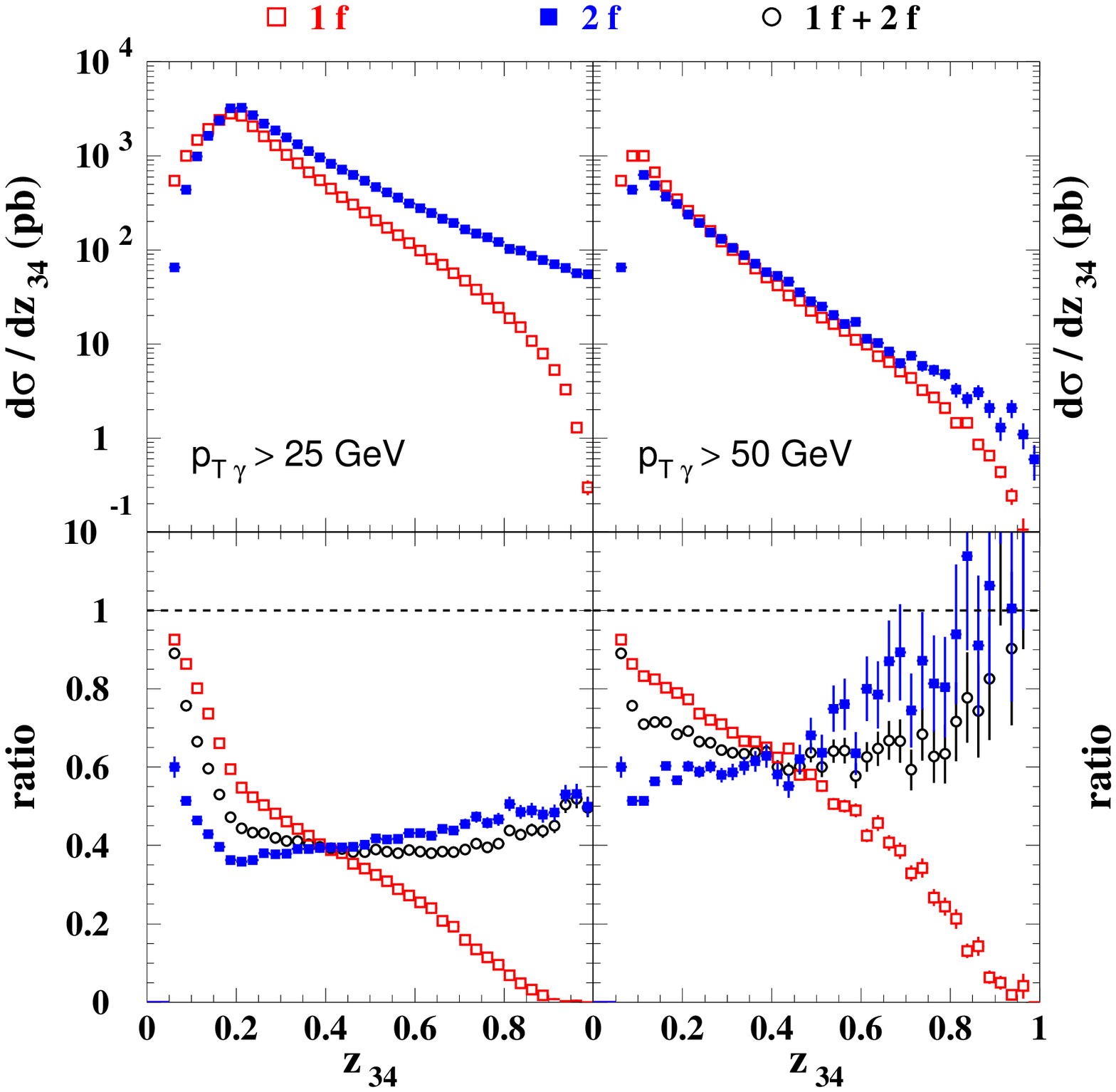}
\end{center}
\vspace{-0.8cm}
\caption{   The 1f and 2f components in \gampi production for various
kinematical configurations for lead-lead scattering at $\sqrt{s} =
5.5$~TeV.  {\it Top:} The $z_{_{3 4}}$ distribution for \ptgamma $>$
25~GeV ({\it left}) and  \ptgamma $>$ 50~GeV ({\it right}).   {\it
Bottom:} The same distributions normalized to the proton-proton case.
}
\label{fig:z_components}
\end{figure}

We stressed in Section~\ref{se:correlations} that the momentum balance
$z_{_{3 4}}$ is closely related to the fragmentation variable $z$ when
the photon is produced directly. Therefore, it may look surprising at
first glance not to observe the suppression becoming more important
when $z_{_{3 4}}$ gets close to one, as the ratio of medium over
vacuum fragmentation functions may suggest in
Figure~\ref{fig:quarktophoton2}. To understand the origin of the flat
behavior of this ratio, the individual 1f and 2f contributions to the
$z_{_{3 4}}$ distribution are represented in
Figure~\ref{fig:z_components} ({\it top}). When the \ptgamma cut is
set to 25~GeV, most of the events (for $z_{_{3 4}} > 0.15$)  actually
come from the double fragmentation process and the relative proportion
of this 2f contribution increases  with $z_{_{3 4}}$. For this 2f
component, however, the connection between the momentum balance and
the fragmentation variable is lost. In particular, the pion transverse
momentum gets larger as $z_{_{3 4}}$ increases, thereby reducing the
medium effects. To be more explicit, we show in the lower panel of
Figure~\ref{fig:z_components} the medium effects on the individual 1f
(open squares) and 2f components (full squares) separately. As
expected, the suppression in the 2f channel becomes less important
when $z_{_{3 4}}$ increases unlike the 1f channel, whose suppression
is reminiscent to the $z$ dependence of the medium over vacuum
fragmentation functions, with a vanishing ratio at $z_{_{3 4}}\simeq
1$. Summing the two contributions, the resulting suppression (open
circles) is an interplay between the 1f and 2f behavior. As far as
counting rates are concerned they are rather large: for ALICE one
expects about 5~10$^5$
 pairs at $z_{_{34}}  = 0.5$ when \ptgamma $>$
25~GeV and  5 10$^4$ 
pairs when \ptgamma $>$ 50~GeV.

In order to isolate the 1f channel --~whose medium effect is
remarkable~-- it would be necessary to increase the photon transverse
momentum, making the 2f process highly unlikely. It can be seen from
the right panel of Figure~\ref{fig:z_components} that going from a
25~GeV to a 50~GeV \ptgamma cut indeed increases significantly the 1f
component. Nevertheless, the 2f contribution remains too large at
large $z_{_{3 4}}$ to observe a huge medium suppression in this
kinematical region. It may then be necessary to trigger on even more
energetic photons, the drawback of too stringent cuts being the
smallness of the corresponding cross sections.

As we shall see in the next Section, diphoton production mostly comes
from the 1f contribution process at the LHC. This observable may
therefore be more interesting than \gampi correlations, at least
regarding the momentum imbalance distributions.

\section{Phenomenology of \gamgam correlations}

\begin{figure}[!ht]
\begin{center}
\includegraphics[height=5.5cm]{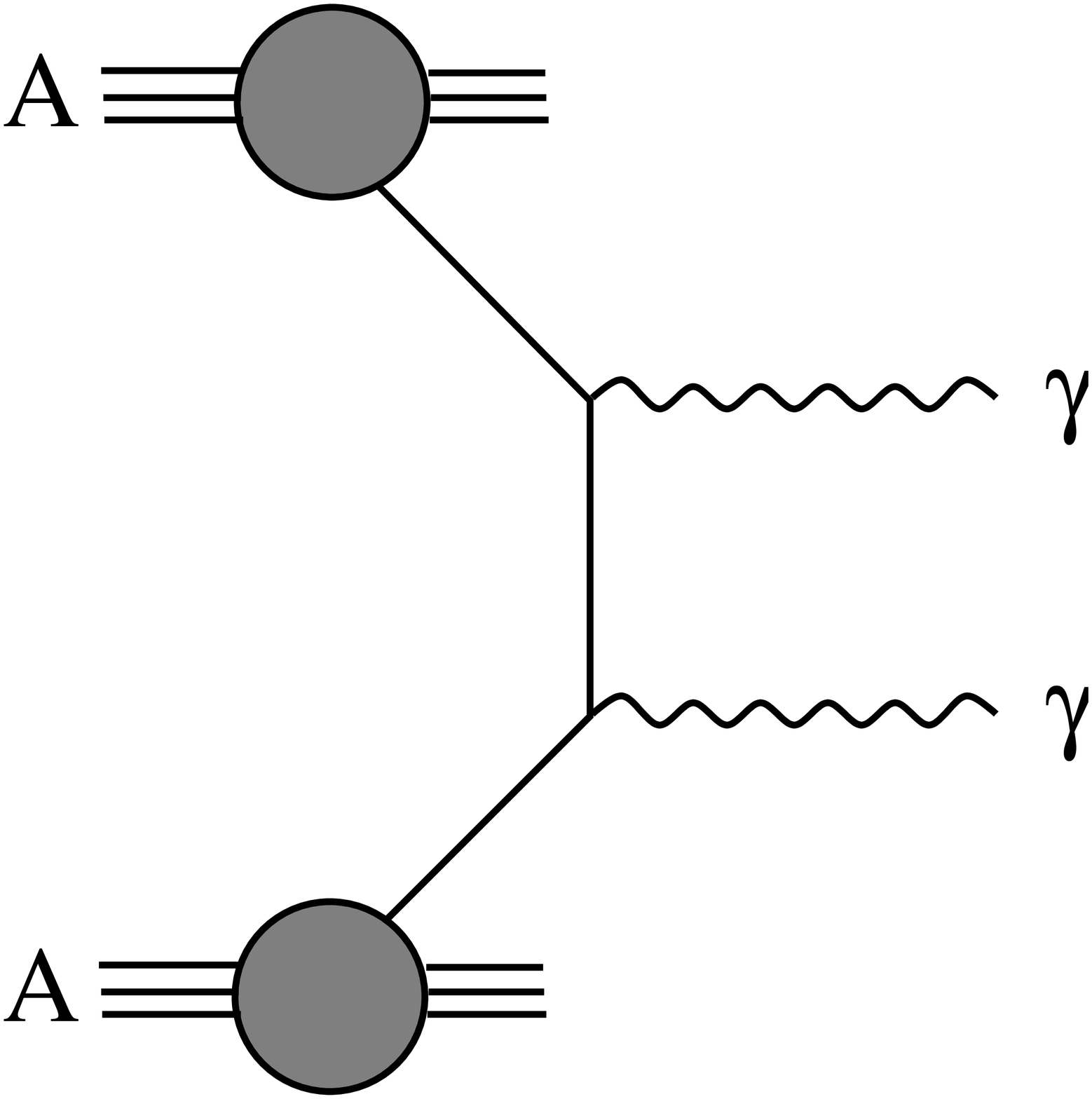}
\end{center}
\vspace{-0.1cm}
\caption{On top of the one-fragmentation and two-fragmentation
process, two photons can be produced directly in \gamgam production at
leading order. This process is not affected by the medium and yields
singular contributions at \qt = 0~GeV and $z_{_{3 4}} = 1$.}
\label{fig:gamgam_direct}\end{figure}  
\begin{figure}[!ht]
\begin{center}
\includegraphics[height=14cm]{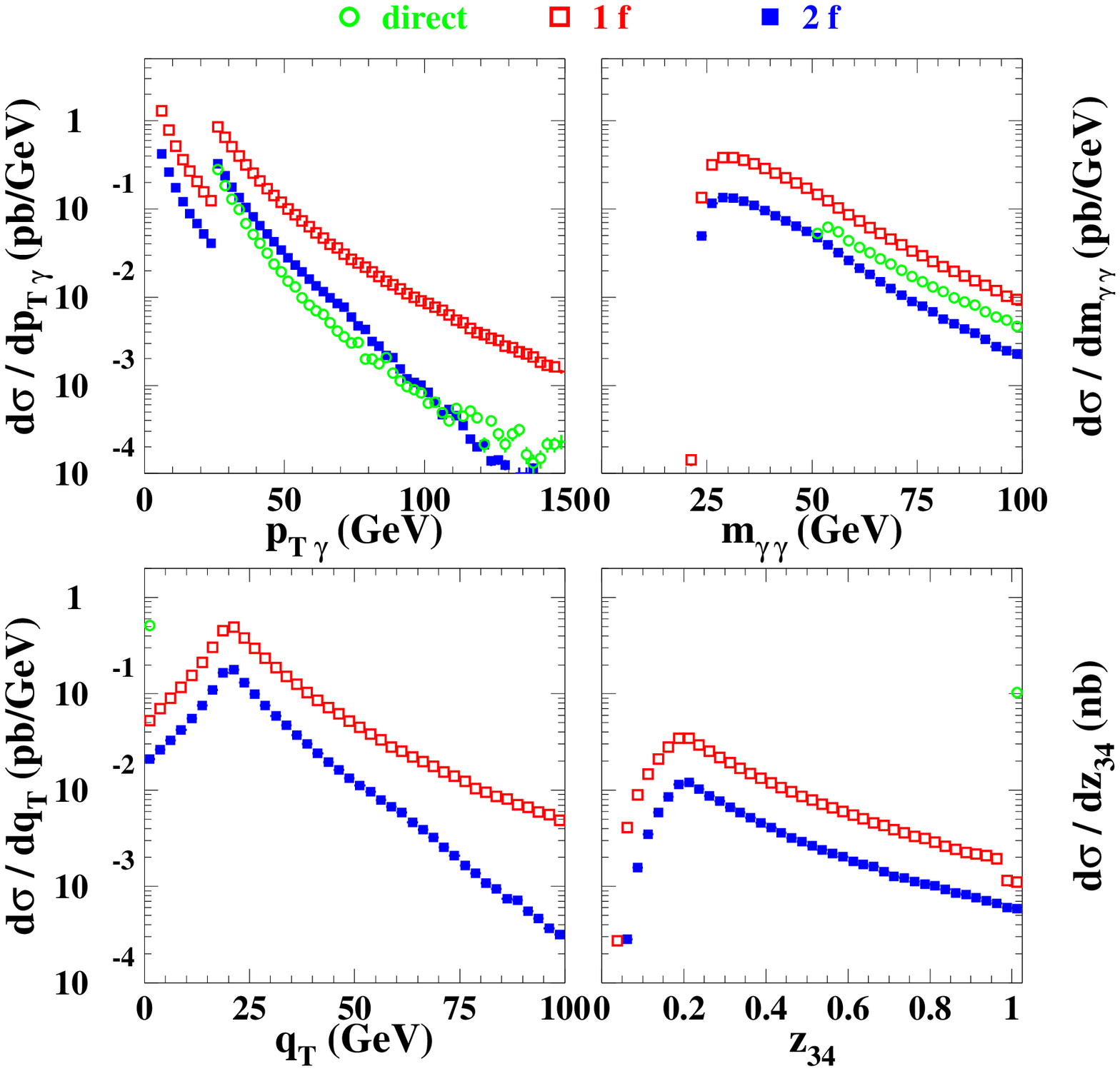}
\end{center}
\vspace{-0.1cm}
\caption{The three components (direct, 1f and 2f) contributing to
diphoton production in proton-proton collisions at $\sqrt s =$~5.5
TeV. A lower cut of 25 GeV is imposed on the transverse momentum of
one photon and 5 GeV on the other. The four distributions shown are in
the transverse momentum of either photon, the diphoton invariant mass,
the transverse momentum of the pair, the diphoton invariant mass and
the momentum imbalance.  }
\label{fig:gamgam_component1}
\end{figure}  


\subsection{Dynamical components}

This section is devoted to the study of \gamgam (or diphoton)
correlations in the same kinematic regime as before. As compared to
the previous cross sections they are, in principle, a factor
$\cO{\alpha/\alpha_s}$ smaller, but the counting rates at the LHC
should nevertheless remain sufficient for our studies. On top of the
1f and 2f components, the new feature is that both photons can be
produced directly ({\it direct} process,
Figure~\ref{fig:gamgam_direct}) in which case they are not affected by
the medium.

The relative weight of all three components, for proton-proton
scattering, are shown in Figure~\ref{fig:gamgam_component1} imposing a
minimum transverse momentum of 25 GeV on one photon and 5 GeV on the
other.  We first consider ({\it top left}) the spectrum in
\ptgamma. It shows a discontinuity at 25 GeV since below this value
only one photon is measured and, furthermore, only the 1f and 2f
processes contribute. One observes the dominance of the 1f component
for the whole transverse momentum range, even when the direct process
contributes, {\em i.e} above  the larger \ptgamma cut.  Concerning the
invariant mass spectrum, the 2f component is at least three times
smaller than the 1f contribution, while the direct piece causes a
small hump to the total cross section at twice the \ptgamma ~threshold
value.  Note the ``singular" contributions of the direct component at
\qt =~0 and $z_{_{34}}$ = 1, a feature of the leading logarithmic
approximation. The ordinate of the corresponding points depend, of
course, on the size of the bin (equivalently the resolution of the
detector). The shape of the distributions near this infrared singular
point is expected to be modified by the higher order corrections.

\subsection{Distributions}

\begin{figure}[!ht]
\begin{center}
\includegraphics[width=16cm]{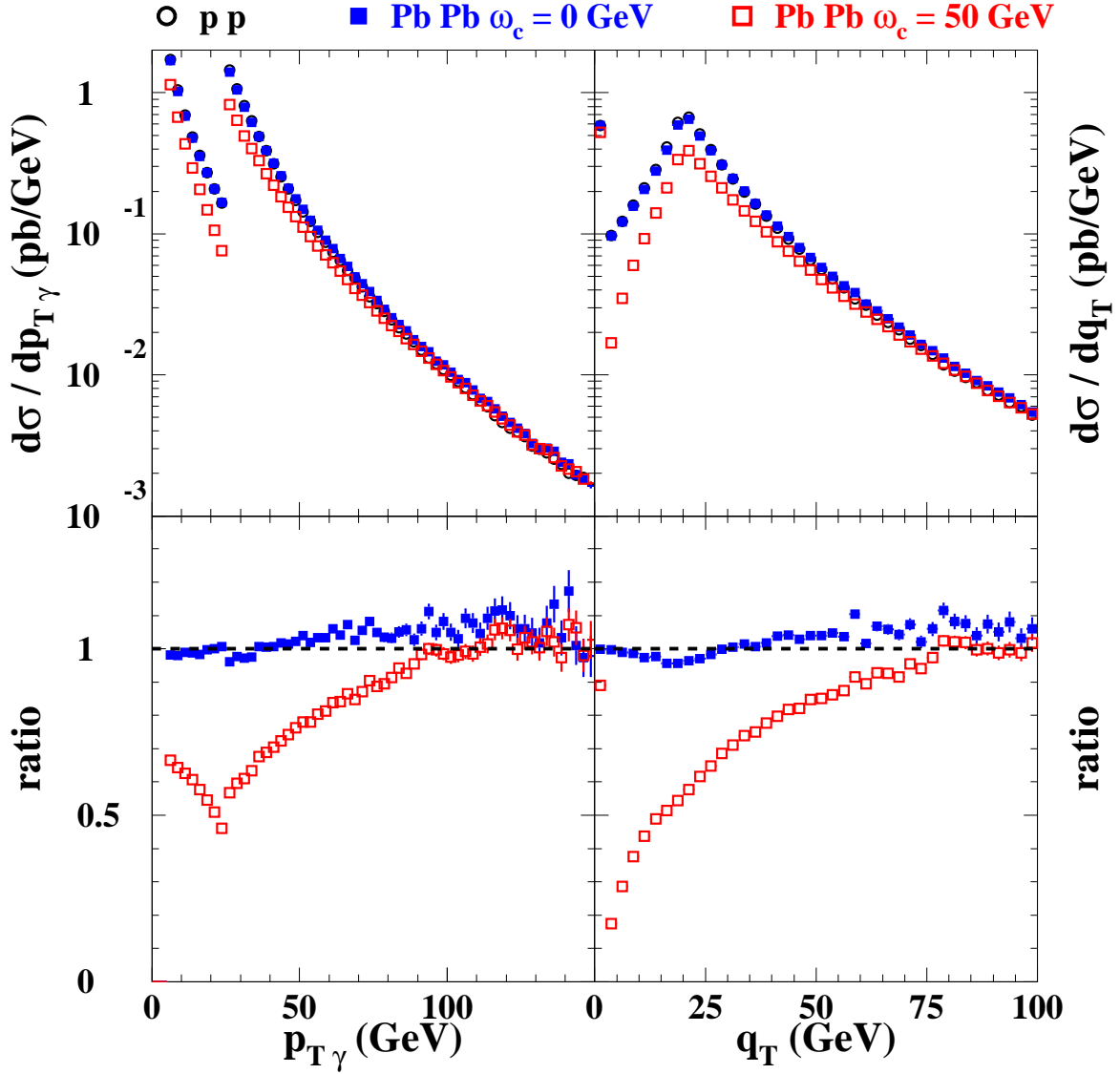}
\end{center}
\vspace{-0.8cm}
\caption{   {\it Top:} The photon \ptgamma and photon pair \qt
transverse momentum distributions in \gamgam production for
proton-proton (open dots) and lead-lead scattering (no energy loss:
black squares; with energy loss: open squares) at $\sqrt{s} =
5.5$~TeV. Both photons are produced at rapidity [-0.5, 0.5] and the
following cuts are imposed: $p_{_T{_{\gamma_1}}} > 25$~GeV and
$p_{_T{_{\gamma_2}}} > 5$~GeV. {\it Bottom:} The same distributions
normalized to the proton-proton case.  }
\label{fig:quarteron_gam_gam_25_05}
\end{figure}

We now compare in Figure~\ref{fig:quarteron_gam_gam_25_05} the
\ptgamma ~and \qt spectra in proton-proton scattering (open dots),
lead-lead scattering with shadowing but without energy loss (full
squares) and lead-lead scattering with shadowing and energy loss (open
squares). The ratios of the lead-lead over proton-proton spectra are
also shown.  Similarly to the \gampi case, the effects of nuclear
shadowing turn out to be negligible. On the contrary, interesting
features due to the energy loss mechanism are observed. The strongest
suppression of the \ptgamma spectra is reached for transverse momenta
of order of the upper cut, \ptgamma$\simeq$~20~GeV. This can be
understood as follows. As $p_{_{T_{\gamma}}}$ approaches the upper cut
from ``below'', events with larger $z$ are selected,
$p_{_{T_{\gamma_2}}} \simeq p_{_{T_{\gamma_1}}}$, where energy loss
effects are most pronounced
(cf. Figure~\ref{fig:quarktophoton}). Above that cut, the proportion
of directly produced photons (unaffected by the medium) is getting
larger and the quenching factor is slowly reaching unity as expected
at asymptotic energies, $p_{_{T_{\gamma}}} \gg \omega_c$. Looking at
the ratio of $q_{_T}$ spectra, the smaller the \qt the larger the
suppression of diphoton events in lead$-$lead collisions.  Indeed,
since the production is dominated by the 1f process, with $z_3 = 1$,
small $q_{_T}$ events imply a large $z_4$ value for the other photon
($q_{_T}=k_{_T}\ |z_3 -z_4|$) where medium effects are the strongest.
Moreover, we no longer observe the same feature as in \gampi
production --~the mild increase of the ratio at very small \qt
(Figure~\ref{fig:quarteron_pi0_gam_25_05}, {\it lower right})~-- since
the 2f fragmentation contribution to diphoton production is much
smaller. Finally, the ratio at $q_{_T} = 0$~GeV is almost close to one
due to the singular contribution of the direct process, unaffected by the medium.

\begin{figure}[!ht]
\begin{center}
\includegraphics[height=14.0cm]{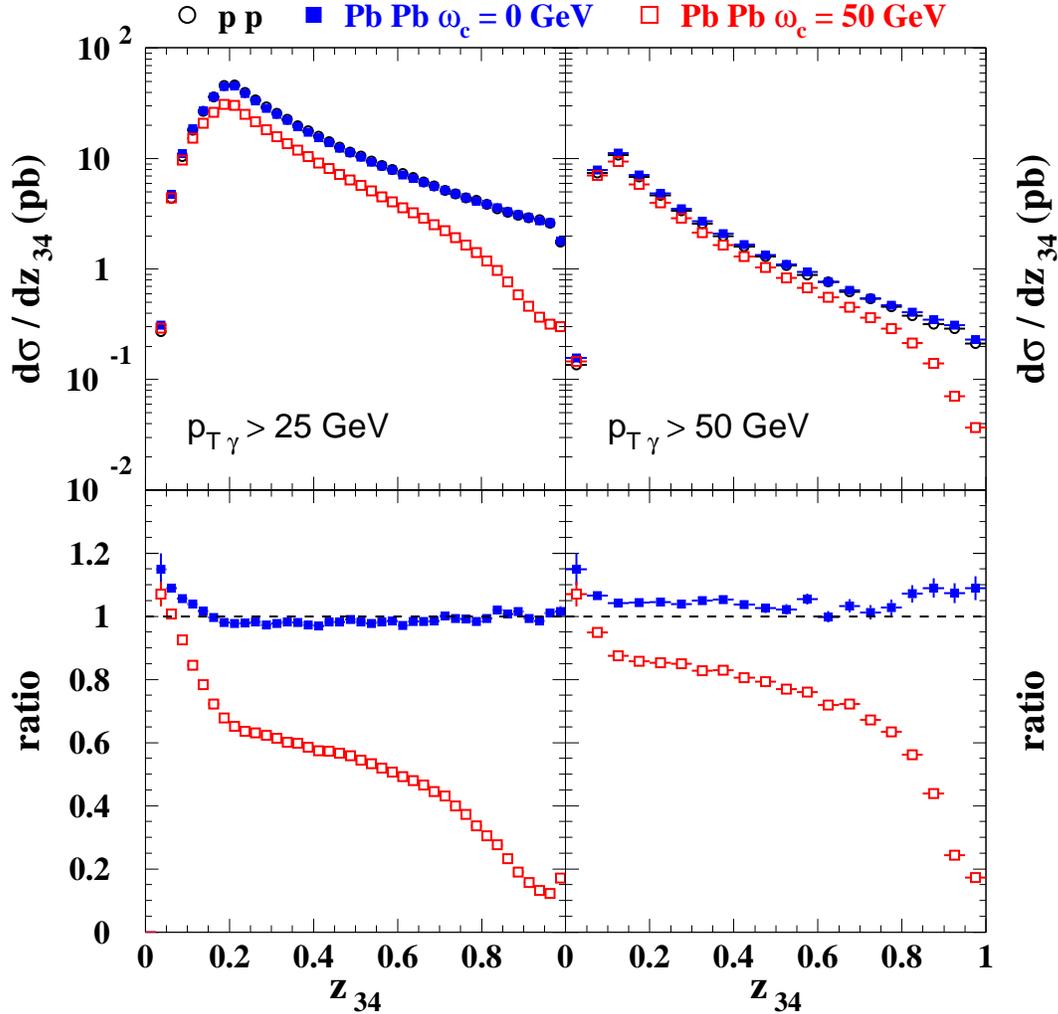}
\end{center}
\vspace{-0.8cm}
\caption{   The $z_{_{3 4}}$ distribution in \gamgam production for
proton-proton (open dots) and lead-lead scattering (no energy loss:
black squares; with energy loss: open squares) at $\sqrt{s} =
5.5$~TeV. Both photons produced at rapidity [-0.5, 0.5] and the
following cuts are imposed: $p_{_T{_{\gamma_1}}} > 25$~GeV and
$p_{_T{_{\gamma_2}}} > 5$~GeV ({\it left}) and $p_{_T{_{\gamma_1}}}
\ge 50$~GeV and $p_{_T{_{\gamma_2}}} \ge 5$~GeV ({\it right}).  {\it
Bottom:} The same distributions normalized to the proton-proton case.
}
\label{fig:z_gam_gam}
\end{figure}

The distribution in the diphoton momentum imbalance $z_{_{3 4}}$ is
now discussed in Figure~\ref{fig:z_gam_gam}, using the previous
kinematical cuts $p_{_{T_{\gamma_1}}} \ge 5$~GeV and
$p_{_{T_{\gamma_1}}} \ge 25$~GeV ({\it left}) as well as
$p_{_{T_{\gamma_2}}} \ge 5$~GeV and $p_{_{T_{\gamma_1}}} \ge 50$~GeV
({\it right}) to keep the parallel with Section~\ref{se:gampi}.
Similarly to the \gampi distribution, the distribution is maximal
around the ratio of the \ptgamma cuts ($z_{_{3 4}} = 0.2$ and 0.1
respectively) and decreases rapidly with $z_{_{3 4}}$, a shape
reminiscent of the photon fragmentation functions in
Figure~\ref{fig:quarktophoton}. In particular, it is remarkable to
notice how the diphoton quenching in Figure~\ref{fig:z_gam_gam} ({\it
bottom}, open squares) ressembles the ratio of medium-modified over
vacuum parton to photon fragmentation functions
(Figure~\ref{fig:quarktophoton2}). Unlike the \gampi case, the smaller
2f contribution to diphoton production does not spoil too much the
large $z_{_{3 4}}$ suppression and make the interpretation of the
momentum imbalance spectra much easier in terms of photon
fragmentation functions.

Notice that one expects a reasonable number of photon-photon events:
for example, for $z_{_{34}} = 0.5$ one has $2~10^4$
 events for \ptgamma
$>$ 25~GeV and $2.5~10^3$  for \ptgamma $>$ 50~GeV with ALICE
luminosity for one month running time.

\section{Qualitative effects of NLO corrections}\label{se:nlo}

\begin{figure}[!ht]
\begin{center}
\includegraphics[height=14.0cm]{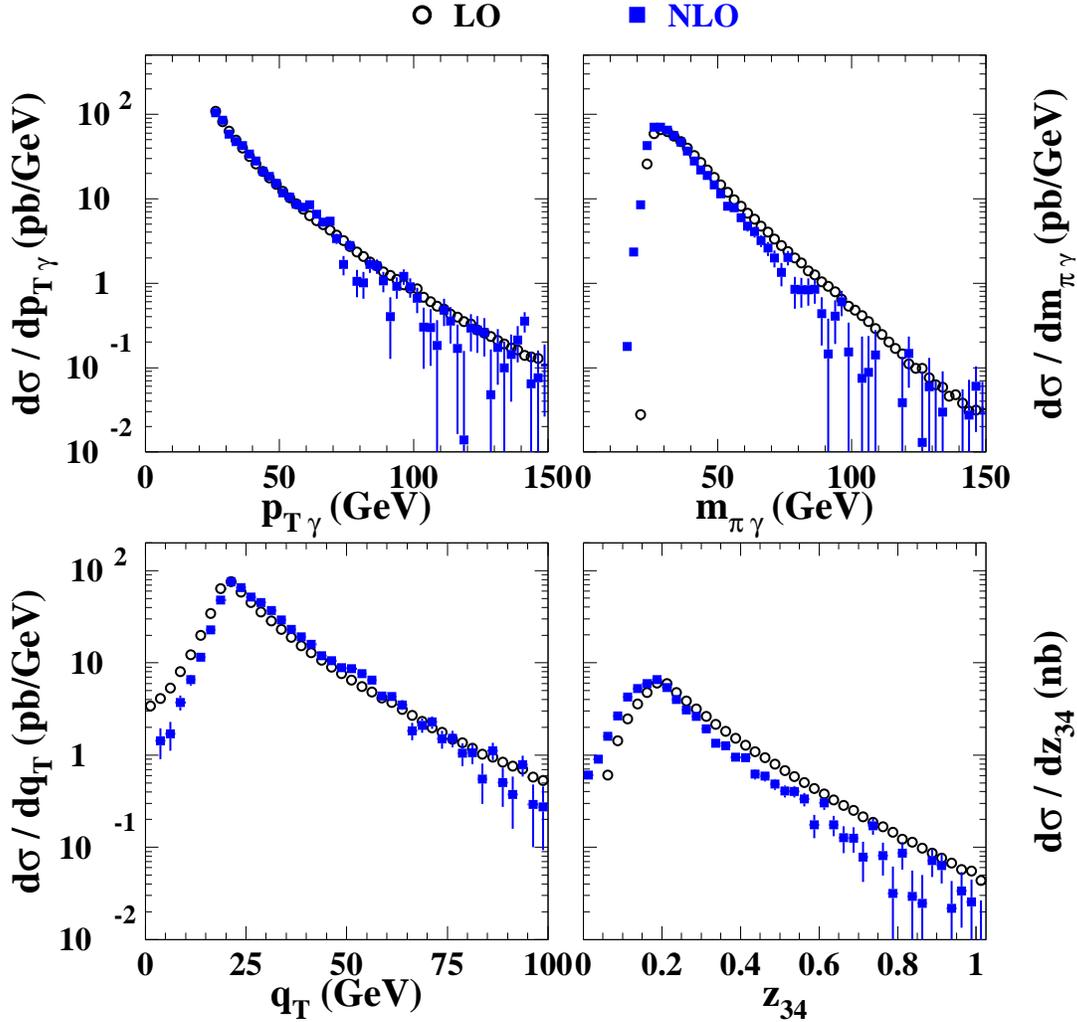}
\end{center}
\vspace{-0.8cm}
\caption{   Comparison of LO and NLO calculations for various
$\pi-\gamma$ correlations.  }
\label{fig:pi0gamNLO}
\end{figure}

\begin{figure}[!ht]
\begin{center}
\includegraphics[height=14.0cm]{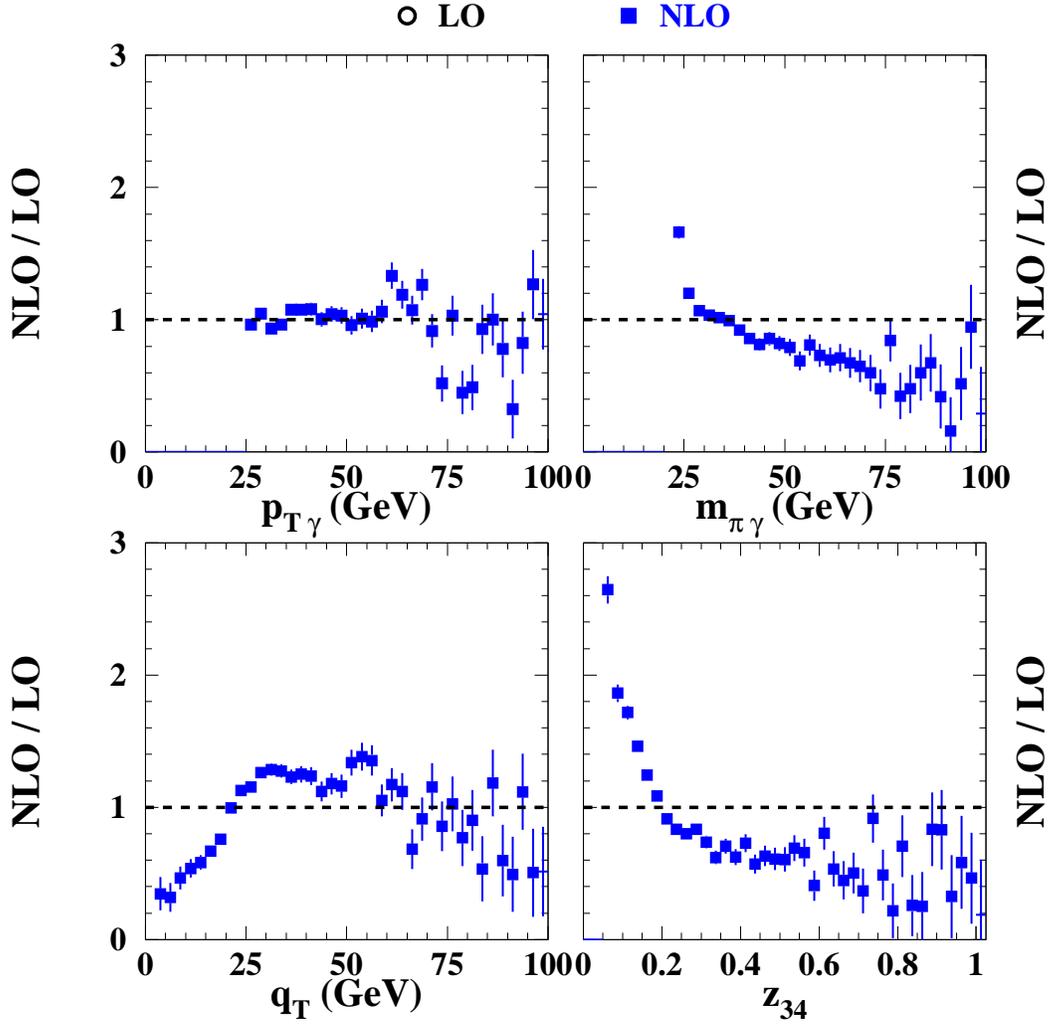}
\end{center}
\vspace{-0.8cm}
\caption{   Ratio of NLO over LO estimates of the $\pi-\gamma$
correlations shown in Figure~\ref{fig:quarteron_pi0_gam_25_05}.  }
\label{fig:pi0gamNLO-ratio}
\end{figure}

\begin{figure}[!ht]
\begin{center}
\includegraphics[height=14.0cm]{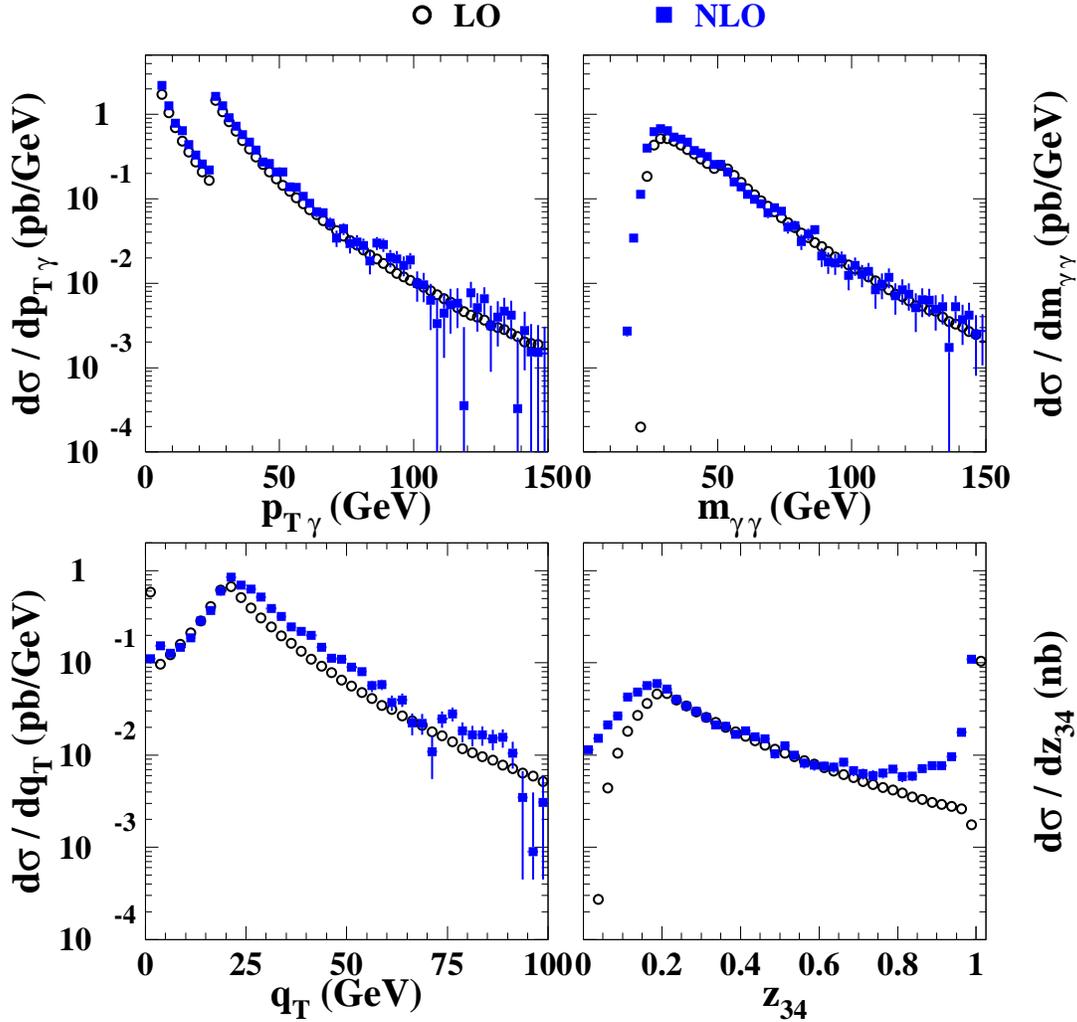}
\end{center}
\vspace{-0.8cm}
\caption{The NLO correlations in \gamgam production for proton-proton
scattering at $\sqrt{s} = 5.5$~TeV.  }
\label{fig:gamgam-NLO-correl}
\end{figure}

\begin{figure}[!ht]
\begin{center}
\includegraphics[height=14.0cm]{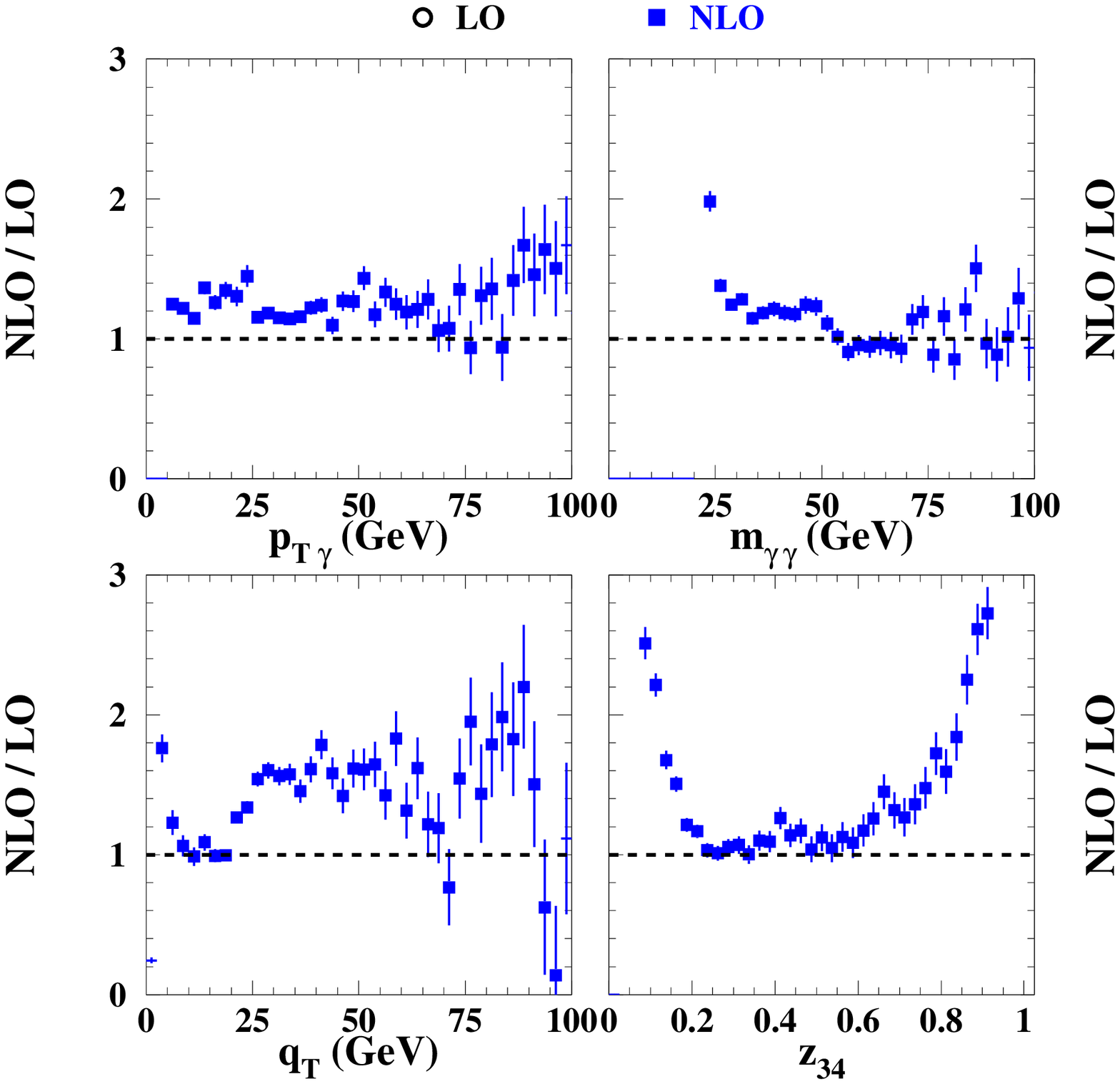}
\end{center}
\vspace{-0.8cm}
\caption{Ratio of NLO correlations over LO correlations for \gamgam
production.  See Figure.~\ref{fig:gamgam-NLO-correl} for details.  }
\label{fig:gamgam_NLO-ratios}
\end{figure}

As already mentioned, the not yet clarified status of NLO QCD
calculations in the medium has lead us to consider \gampi and \gamgam
production to leading order accuracy. Nevertheless,  it is instructive
to study and quantify the role of higher order corrections in
proton-proton collisions. In particular, attention should be paid to
the phase space which gets modified by higher order QCD corrections:
consequently this can affect our leading order predictions in
lead-lead collisions.

The NLO predictions are obtained using the CTEQ6M structure
functions~\cite{cteq6}, the KKP next-to-leading logarithmic
fragmentation functions for the $\pi^0$ and the BFG, set II,
fragmentation functions into a photon. As before all scales are set
equal to ($p_{_{T_3}}$ + $p_{_{T4}}$)/2. A cut in the azimuthal angle
between the two particles has been applied, $\phi \ge \phi_{\rm{min}}
= \pi/2$.

The \gampi correlation functions are plotted in
Figure~\ref{fig:pi0gamNLO} at LO ({\it dots}) and NLO ({\it
squares}). The overall effect of higher order corrections proves quite
small --~say less than 30 \%~-- except in some specific regions of
phase space. As shown in Figure~\ref{fig:pi0gamNLO-ratio}, the ratio
of NLO over LO spectra is almost constant except at small invariant
mass or momentum imbalance. The reason comes from the new parton
configurations in momentum space available at higher order. Indeed,
the two particles are no longer constrained to have opposite momenta
when going from the two-body to the three-body NLO kinematics. This
will affect, in particular, variables like the invariant mass, the
pair transverse momentum or the momentum imbalance which now depend
explicitely on the relative azimuthal angle $\phi$ between the two
particles  (see Eqs. (\ref{eq:invmass}) to (\ref{eq:scaledp34})). This
can be seen in Figure~\ref{fig:pi0gamNLO-ratio} where the momentum
imbalance spectrum is clearly enhanced at NLO when $z_{34} \propto
\cos{\phi}$ gets very small: this corresponds to configurations where
the observed particles are recoiling from the third undetected jet in
the opposite hemispere.  Other effects include the shift of the
threshold in the invariant mass distribution, from $m_{_{\pi\gamma}} = \sqrt{4
p_{_{T_\gamma}} p_{_{T_\pi}}} = 10\sqrt{5}$~GeV ($\phi=\pi$) to
$m_{_{\pi\gamma}} = \sqrt{2 p_{_{T_\gamma}} p_{_{T_\pi}}} = 5\sqrt{10}$~GeV
($\phi=\pi/2$), or the shift of the pair momentum spectrum to larger
\qt which results in the ratio smaller than one below 20~GeV (the
difference of the $p_{_T}$ cuts, i.e. the maximum of the distribution)
and larger above. All these effects depend crucially on the cut
$\phi_{\rm{min}}$ in the azimuthal angle and should vanish as
$\phi_{\rm{min}}$ approaches $\pi$.

The \gamgam correlation functions displayed in
Figure~\ref{fig:gamgam-NLO-correl} indicate that higher order
corrections do not strongly modify the LO results in this channel
either, except near the infrared singular point ($q_{_T} = 0$~GeV or
$z_{_{34}}=1$) or in the domain where new phase space is available
(small $m_{_{\gamma\gamma}}$ or small $z_{_{34}}$). In the latter case
the NLO momentum configurations modify the LO spectra in a way similar
to the \gampi correlations (see the spectrum ratios in
Figure~\ref{fig:gamgam_NLO-ratios}). Although the effect proves tiny,
we may also remark the lower threshold for the direct process in the
transverse momentum distribution, now slightly below the 25~GeV
cut. However, the most remarkable feature when going from LO to NLO in
\gamgam production deals with the infrared sensitivity of observables
such as the transverse momentum or the momentum imbalance
spectrum. When \qt gets small as compared to the diphoton invariant
mass, the phase space restriction forces the emitted gluons to be
extremely soft. The $\delta$ function singularity which appeared in
the leading logarithm approximation now spreads in phase space, due to
the partial cancellation of real and virtual NLO terms, and it is
broadened at NLO accuracy. Indeed, we notice in
Figure~\ref{fig:gamgam-NLO-correl} the significant corrections which
extend up to roughly \qt$\lesssim 10$~GeV. This gives us a typical
range in which the present perturbative calculation may not be
reliable. Since $q_{_T}/m_{\gamma\gamma}\propto 1-z_{_{34}}$, such a
behavior can also be observed in the momentum imbalance spectrum near
the singular point, $z_{34}=1$ where the NLO results start to deviate
significantly from the LO prediction above $z_{34}\gtrsim 0.8$.
Technically, large terms such as
$\alpha_s\,\ln^2\left(q^2_{_T}/m^2_{\gamma\gamma}\right)$  and
$\alpha_s\,\ln\left(q^2_{_T}/m^2_{\gamma\gamma}\right)$ contribute to
the direct process making the present fixed order QCD calculation not
reliable very near the infrared singular point. For a more accurate
approximation such large terms should be resummed. Although the one
(two) fragmentation component requires one (two) integration(s) over
the scaling variable $z_{_3}$ ($z_{_3}$ and $z_{_4}$), which smears out
these large logarithms~\cite{bgpw2001} and make the
distributions regular at small $q_{_T}$, resummation may affect the shape
of the distributions in this region.

Let us now discuss the phenomenological implications of higher order
corrections to our predictions for the spectrum ratios in lead-lead
over those in proton-proton collisions. First the moderate higher
order corrections (except in specific domains for some correlations)
clearly indicate that neither the absolute attenuation nor the shape
of the attenuation in lead-lead collision should be affected too
much. In fact, the larger presence of gluons at higher order, whose
energy loss is stronger than for quarks, should be responsible for a
slightly more pronounced suppression. In the large \qt or small
$z_{_{34}}$ region, where the NLO over LO ratio is the largest due to
the non-collinear configurations, one can also expect the quenching to
prove more pronounced. Indeed, such regions were not affected much by
the medium in our LO prediction as they require the fragmentation of
very large $\kt$ partons, hence with a small energy loss effect. At
NLO, however, the large \qt and small $z_{_{34}}$ domain can be
reached while keeping the parton energy $\kt$ not too large (as
compared to $\omega_c$), provided the relative azimuthal angle between
the two particles is small enough.

We emphasized in the previous section the strong attenuation of
diphoton production expected in lead-lead collision near the boundary
of phase space, in particular at large $z_{34}$ and small $q_{_T}$. On
the other hand, the presence of the direct process, unaffected by
parton energy loss, should make the ratio equal to one exactly at
$z_{34}=1$ and \qt=~0~GeV. The competition between the direct and the
fragmentation process at LO therefore generated discontinuities in the
ratio at these specific points (see
e.g. Figure~\ref{fig:quarteron_gam_gam_25_05}), which should be
smoothed at higher order. Based on the present NLO calculation in
proton-proton reactions, we expect that the quenching of diphoton
production should start to increase below \qt$\lesssim 10$~GeV or
$z_{34}\gtrsim 0.8$. Similarly, the discontinuity seen in
Figure~\ref{fig:quarteron_gam_gam_25_05} for 25~GeV photon transverse
momenta should be smeared as well.

\section{Background}

\begin{figure}[!ht]
\begin{center}
\includegraphics[height=14.0cm]{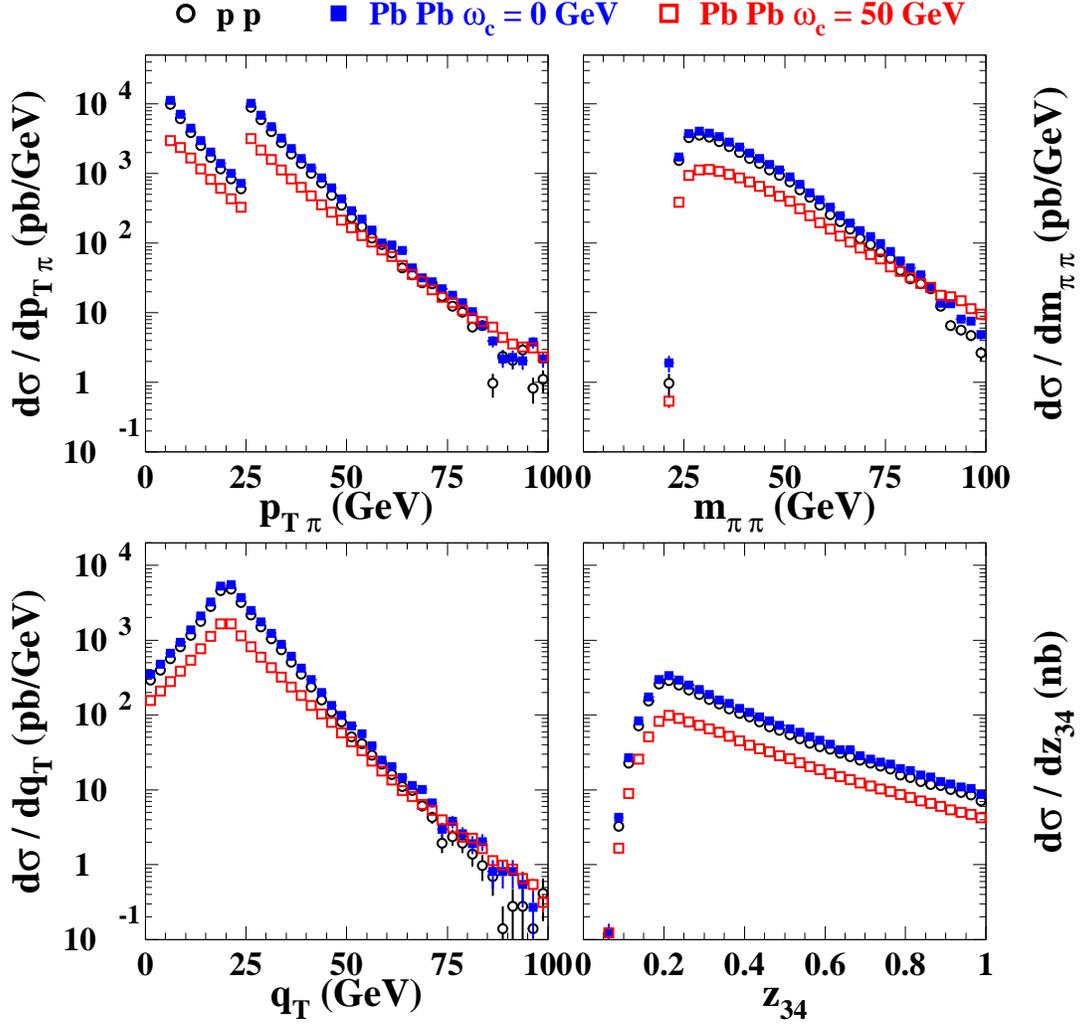}
\end{center}
\vspace{-0.8cm}
\caption{ \pipi correlations. Both pions are produced at rapidity
[-0.5, 0.5] and the following cuts are imposed: $p_{_T{_{\pi_1}}} >
25$~GeV and $p_{_T{_{\pi_2}}} > 5$~GeV. The conventions for the
symbols are as in Figure~\ref{fig:quarteron_pi0_gam_25_05}.  }
\label{fig:pipi}
\end{figure}

In the photon transverse momentum range discussed above the background
from $\pi^0$ decays will still be appreciable. In order to illustrate
this background we briefly present here various \pipi distributions,
using the same asymmetrical cuts as before, namely
$p_{_{T_{\pi_1}}}>25$~GeV and $p_{_{T_{\pi_2}}}>5$~GeV (see
Figure~\ref{fig:pipi}). In this case only the 2f mechanism contributes
and, as a consequence, the distributions display very similar features
to the \gampi case. Only the size of the correlations is larger by
roughly a factor 50. Such distributions should therefore be determined
with a great accuracy to be subtracted in order to measure the \gampi
and \gamgam distributions discussed so far.

Although one may still notice significant effects in nucleus-nucleus
collisions, medium-modified fragmentation functions cannot be
determined through $\pi^0-\pi^0$ correlations in the absence of the 1f
component. Nevertheless, we feel it should be interesting to perform
such correlations with low \ptpi~cuts for {\it both} pions which would
possibly allow to study the spatial distribution of the hot
medium. This has been achieved for instance by the RHIC experiments
who considered the \pipi azimuthal
correlations~\cite{rhic}. Theoretically, this would  require a
complete description of the space-time energy density though
--~available e.g. in hydrodynamical models~-- which go beyond the
scope of the present study. Such attempts have been suggested
recently~\cite{Eskola:2004cr,Dainese:2004te}.

\section{Conclusions}

We have discussed various photon tagged correlations as a tool to
study jet fragmentation in the hot medium created in heavy-ion
collisions.  Correlations functions have been computed to leading
order in proton-proton collisions at LHC energy. Similar distributions
were determined in lead-lead collisions, assuming medium-modified
fragmentation functions to account for the parton energy loss process
in the dense medium.

We show that significant effects could be expected at LHC energy both
in the \gampi and \gamgam channel. Ideally, the use of asymmetric cuts
in the transverse momentum of both particles  allow the possibility to
map out the parton fragmentation functions modified by the
medium. However several production mechanisms co-exist with a relative
weight varying with the kinematical variable under consideration. This
smears somehow the relation between observables and the fragmentation
functions. Consequently, we found various suppression patterns,
depending on the imposed kinematical constraints, which should be
accessible experimentally. Furthermore, the variety of observables
presented here should help to constrain the underlying model for
parton energy loss.

Although calculations were performed at leading order, NLO corrections
have also been addressed. In particular, the way higher order
corrections could modify the expected quenching of \gampi and \gamgam
spectra is discussed. To be more specific, we believe our present LO
prediction to be reliable up to roughly $z_{_{34}}\simeq
0.8$. Finally, \pipi correlation functions were computed so as to give
a reference for the expected background one could face at the LHC
within the kinematic cuts we employed.

\section*{Acknowledgments}

The authors thank Hugues Delagrange and Monique Werlen for discussions.

\end{document}